\pdfoutput=1
\documentclass[12pt,a4paper,final]{iopart}
\usepackage[latin1]{inputenc}
\usepackage[english]{babel}
\usepackage{iopams,setstack,enumerate}
\usepackage{graphicx}
\usepackage{stmaryrd}
\usepackage[title]{appendix}
\SetSymbolFont{stmry}{bold}{U}{stmry}{m}{n}
\usepackage[breaklinks=true,colorlinks=true,linkcolor=blue,urlcolor=blue,citecolor=blue]{hyperref} 

\newtheorem{prop}{Proposition}[section]
\eqnobysec

\begin{document}

\title[Geometry of the basic statistical physics mapping]{Geometry of the basic statistical physics mapping}

\author{Mario Angelelli$^{1}$
and Boris Konopelchenko$^{1}$
}
\address{$^1$Department of Mathematics and Physics ``Ennio De Giorgi'', University of Salento and Sezione INFN, 73100, Lecce, Italy }
\eads{\mailto{mario.angelelli@le.infn.it}, \mailto{boris.konopeltchenko@unisalento.it}}

\begin{abstract}

Geometry of hypersurfaces defined by the relation which generalizes classical formula for free energy in terms of microstates is studied. Induced metric, Riemann curvature tensor, Gauss-Kronecker curvature and associated entropy are calculated. Special class of ideal statistical hypersurfaces is analyzed in details. Non-ideal hypersurfaces and their singularities similar to those of the phase transitions are considered. Tropical limit of statistical hypersurfaces and double scaling tropical limit are discussed too. 

\end{abstract}

\vspace{2pc}


\section{Introduction}

One of main formulae of statistical physics 

\begin{equation}
{\displaystyle F=-kT\ln\left(\sum_{\{n\}}e^{-{\textstyle \frac{E_{\{n\}}}{kT}}}\right)}\label{eq: free energy}
\end{equation}
establishes the relation between the set of microstates of the macroscopic
systems with the energies $E_{\{n\}}$, enumerated by quantum numbers
$\{n\}$, at the temperature $T$ ($k$ is the Boltzmann constant),
and the macroscopic free energy $F$ (see e.g. \cite{LL1980}). The microstates are realized with the
Gibbs probability 

\begin{equation}
{\displaystyle w_{\{n\}}=\frac{\exp\left(-\frac{E_{\{n\}}}{kT}\right)}{{\displaystyle \sum_{\{m\}}}\exp\left(-\frac{E_{\{m\}}}{kT}\right)}}.\label{eq: Gibbs distribution}
\end{equation}

Mathematically, the formula (\ref{eq: free energy}) provides us with
the mapping of the point-set $\{f_{1},f_{2},\dots\}$ to the function
${\displaystyle f=\ln\left(\sum_{\{n\}}e^{f_{\{n\}}}\right)}$ in
terms of the quantities ${\displaystyle f_{\{n\}}\equiv-\frac{E_{\{n\}}}{kT}}$
and ${\displaystyle f\equiv-\frac{F}{kT}}$. In such a micro-macro
mapping these quantities depend, in fact, on
several parameters. First, they are explicitly functions of the temperature
$T$. Then, the energy spectrum $E_{\{n\}}$ typically depends on
some parameters. In addition, the standard approach to complex systems
requires to deal with a family of systems with close properties (e.g.,
with varying $E_{\{n\}}$) in order to reveal their characteristic
features. 

All this suggests to consider, instead of the mapping (\ref{eq: free energy}),
the general mapping 

\begin{equation}
F=\ln\left(\sum_{\alpha=1}^{m}e^{f_{\alpha}(x_{1},\dots,x_{n})}\right)\label{eq: basic statistical mapping}
\end{equation}
with real-valued functions $f_{\alpha}$ of $n$ real variables $x_{1},\dots,x_{n}$
and arbitrary $n$ and $m$. Study of geometrical objects related
to the mapping (\ref{eq: basic statistical mapping}) and probabilities 

\begin{equation}
w_{\alpha}=\frac{e^{f_{\alpha}(\boldsymbol{x})}}{{\displaystyle \sum_{\beta=1}^{m}e^{f_{\beta}(\boldsymbol{x})}}},\quad\alpha=1,\dots,m\label{eq: basic Gibbs probabilities}
\end{equation}
is the main goal of this paper. 

Geometrical structures associated with the standard thermodynamics
have been already discussed many times (see e.g. \cite{Rao1945,Mrugala1988,Gilmore1984,QSTV2010,QSTV2011,QQ2014,BL-MN2015,Passare2012}).
Interrelation of geometry and certain statistical models has been
considered too (see e.g. \cite{Amari2000,Watanabe2009,Sun2014,Nie2014,Dodson2012}). 

In the present paper we, in contrast to the phenomenological Geometrothermodynamics
\cite{QSTV2011,QQ2014}, start with the generalization (\ref{eq: basic statistical mapping})
of the micro-macro mapping (\ref{eq: free energy}). We
treat it as the definition of the $n$-dimensional
hypersurface $V_{n}$ (referred hereafter as statistical hypersurface)
in the Euclidean space $\mathbb{R}^{n+1}$ with local coordinates
$(x_{1},\dots,x_{n},x_{n+1}\equiv F)$ and analyse its geometry. Since
${\displaystyle \frac{\partial F}{\partial f_{\alpha}}=w_{\alpha}}$
and ${\displaystyle dF=\sum_{i=1}^{n}\sum_{\alpha=1}^{m}w_{\alpha}\frac{\partial f_{\alpha}}{\partial x_{i}}dx_{i}}$,
the probabilities (\ref{eq: basic Gibbs probabilities}) show up in
all geometrical objects associated with the hypersurface $V_{n}$.
It is a characteristic feature of the hypersurface $V_{n}$ defined
by the formula (\ref{eq: basic statistical mapping}). In particular,
the induced metric $g_{ik}$ is of the form 

\begin{equation}
g_{ik}=\delta_{ik}+\bar{f_{i}}\cdot\bar{f_{k}},\quad i,\,k=1,\dots,n\label{eq: basic induced metric}
\end{equation}
where ${\displaystyle \bar{f_{i}}\doteq\sum_{\alpha=1}^{m}w_{\alpha}\frac{\partial f_{\alpha}}{\partial x_{i}}}$.
Second fundamental form, Christoffel symbols, Riemann curvature tensor and Gauss-Kronecker
curvature are also expressed via this and similar mean values. In general,
the metric (\ref{eq: basic induced metric}) is not flat. 

In addition to the standard geometrical questions, physics suggests
to address the problems which mimic those typical for statistical
physics, for instance properties of ideal and non-ideal systems, phase
transitions \textit{etc.} \cite{LL1980}. The simplest model
of an ideal gas corresponds to linear functions ${\displaystyle f_{\alpha}=\sum_{i=1}^{n}a_{\alpha i}x_{i}+b_{\alpha}}$,
$\alpha=1,\dots,m$, where $a_{\alpha i}$ and $b_{\alpha}$ are constants.
In this case ${\displaystyle \bar{f}_{i}=\sum_{\alpha=1}^{m}w_{\alpha}a_{\alpha i}}$
and all formulae are simplified drastically. However, the Riemann
curvature remains nonvanishing. The special feature of this case is
that the Gauss-Kronecker curvature of such ideal hypersurface $V_{n}$
is equal to zero if rank of the matrix $\boldsymbol{a}=\left(a_{\alpha i}\right)$
is smaller than $n$.

Super-ideal case with $m=n$ and $f_{\alpha}=x_{\alpha}$, $\alpha=1,\dots,n$,
is studied in details. It is shown that scalar and mean curvatures
of the corresponding statistical hypersurfaces are non-negative and
have upper bounds depending on $n$. 

It is shown that for nonlinear functions $f_{\alpha}(\boldsymbol{x})$,
which correspond to non-ideal macroscopic systems even in the simplest
cases $n=2,3$, the Gauss-Kronecker curvature, in general, is different
from zero. 

Phase transitions of the first order for macroscopic systems are mimicked
by discontinuities of the metric (\ref{eq: basic induced metric})
and singularities of the curvature. It is shown that the geometrical
characteristics may have various types of behaviours for such singularities. 

The geometrical interpretation of the classical entropy ${\displaystyle S=-\sum_{\alpha=1}^{m}w_{\alpha}\ln w_{\alpha}}$
and its connection with the coupling between the hypersurface $V_{n}$ and its normal bundle
is considered. It is shown that in the particular case when all functions
$f_{\alpha}(\boldsymbol{x})$ are homogeneous functions of degree
one the entropy is given by the scalar product 
\begin{equation}
S=\sqrt{\det \boldsymbol{g}}\,\overrightarrow{X}\cdot\overrightarrow{N}\label{eq: entropy and scalar product}
\end{equation}
where $\overrightarrow{X}$ and $\overrightarrow{N}$ are position
vector and normal vector at the point on the hypersurface $V_{n}$,
respectively. 

Tropical limit of statistical hypersurfaces is discussed
too. The standard tropical limit of the ideal hypersurface is given
by a piecewise hyperplane. It is shown that the analysis of non-ideal
cases, in general, requires to use double-scaling tropical limit. Piecewise curved
hypersurfaces represent tropical limit of non-ideal hypersurface.
So, in the tropical limit the difference between ideal and non-ideal
cases becomes easily visible. 

The paper is organized as follows. In section (\ref{sec: Geometric characteristics of hypersurface })
formulae for metric, second fundamental form, Riemann curvature tensor and Gauss-Kronecker
curvature of statistical hypersurface $V_{n}$ are presented. Ideal
hypersurfaces which mimic ideal gas are studied in section (\ref{sec: Ideal statistical hypersurfaces }).
Super-ideal hypersurfaces are discussed in section (\ref{sec: Special ideal hypersurfaces}).
Non-ideal cases with nonlinear functions $\{f_{\alpha}\}$ are considered
in section (\ref{sec: Non-ideal case}). Singularities of statistical
hypersurface are analyzed in section (\ref{sec: Singularities of the statistical hypersurface }).
Section (\ref{sec:Tropical limit of statistical hypersurfaces}) is
devoted to the study of tropical limit of statistical hypersurfaces.
Double scaling tropical limit is discussed in section (\ref{sec: Double scaling tropical limit in non-ideal case}).

\section{\label{sec: Geometric characteristics of hypersurface } Geometric
characteristics of statistical hypersurfaces}

The formula (\ref{eq: basic statistical mapping}) can be viewed in
various ways to define geometrical objects. We will follow the simplest
and standard one, i.e. to view the graph of function given by (\ref{eq: basic statistical mapping})
as a hypersurface in $(n+1)$-dimensional Euclidean space with cartesian
coordinates $x_{1},\,x_{2}\dots,\,x_{n},\,x_{n+1}\equiv F$. Thus,
the induced metric of this hypersurface $V_{n}$ is (see e.g. \cite{Eisenhart1997}) 

\begin{equation}
g_{ik}=\delta_{ik}+\frac{\partial F}{\partial x_{i}}\cdot\frac{\partial F}{\partial x_{k}},\quad i,\,k=1,\dots,n.\label{eq: basic metric 1}
\end{equation}
Since 
\begin{equation}
\frac{\partial F}{\partial f_{\alpha}}=\frac{e^{f_{\alpha}(\boldsymbol{x})}}{{\displaystyle \sum_{\beta=1}^{m}e^{f_{\beta}(\boldsymbol{x})}}}\equiv w_{\alpha},\quad \alpha=1,\dots,m,\label{eq: basic Gibbs weights 1}
\end{equation}
one has 
\begin{equation}
g_{ik}=\delta_{ik}+\bar{f}_{i}\cdot\bar{f}_{k},\quad i,\,k=1,\dots,n\label{eq: basic Gibbs weights 2}
\end{equation}
where 
\begin{equation}
\bar{f}_{i}\doteq\sum_{\alpha=1}^{m}w_{\alpha}\frac{\partial f_{\alpha}}{\partial x_{i}},\quad i=1,\dots,n.\label{eq: averages derivative}
\end{equation}
Then $0\leq w_{\alpha}\leq1$, $\alpha=1,\dots,m$ and ${\displaystyle \sum_{\alpha=1}^{m}w_{\alpha}=1}$.
So the quantity $w_{\alpha}$ represents the probability to have function
$f_{\alpha}$ from the set $\{f_{1},f_{2},\dots f_{m}\}$. It is a
membership function in terminology of Fuzzy sets (see e.g. \cite{Zadeh1965,Klir1988}).
The presence of this generalized Gibbs probability (or Gibbs membership
function) and mean values $\bar{f}_{i}$ (\ref{eq: averages derivative})
is a characterizing feature of all geometric quantities associated
with the statistical hypersurface $V_{n}$. 

Using the standard formulae (see e.g. \cite{Eisenhart1997}), one calculates other
characteristics of the hypersurface $V_{n}$. The position vector
$\vec{X}(\boldsymbol{x})$ for a point on $V_{n}$ and the corresponding
normal vector $\vec{N}(\boldsymbol{x})$ are 
\begin{eqnarray}
\vec{X}=(x_{1},\dots,x_{n},x_{n+1}),\nonumber\\
{\displaystyle \vec{N}=\frac{1}{\sqrt{\det\boldsymbol{g}}}\left(-\bar{f}_{1},\dots,-\bar{f}_{n},1\right)}
\label{eq: position vector and normal vector}
\end{eqnarray}
where $\displaystyle
\det\boldsymbol{g}=1+\sum_{i=1}^{n}\bar{f}_{i}^{2}$. 
So the second fundamental form $\Omega_{ik}$ is given by 
\begin{equation}
\Omega_{ik}=\vec{N}\cdot\frac{\partial^{2}\vec{X}}{\partial x_{i}\partial x_{k}}=\frac{1}{\sqrt{\det\boldsymbol{g}}}\left(\bar{f}_{\{ik\}}-\bar{f}_{i}\cdot\bar{f}_{k}\right),\quad i,\,k=1,\dots,n\label{eq: second fundamental form}
\end{equation}
where 
\begin{equation}
\bar{f}_{\{ik\}}\doteq\sum_{\alpha=1}^{m}w_{\alpha}\left(\frac{\partial^{2}f_{\alpha}}{\partial x_{i}\partial x_{k}}+\frac{\partial f_{\alpha}}{\partial x_{i}}\cdot\frac{\partial f_{\alpha}}{\partial x_{k}}\right),\quad i,\,k=1,\dots,n.\label{eq: second average}
\end{equation}
Then, since 
\begin{equation}
{\displaystyle \frac{\partial g_{ik}}{\partial x_{l}}=\sqrt{\det\boldsymbol{g}}\left(\Omega_{il}\bar{f}_{k}+\Omega_{kl}\bar{f}_{i}\right)},\quad l=1,\dots,n\label{eq: derivative metric}
\end{equation}
and 
\begin{equation}
\frac{\partial}{\partial x_{i}}\det\boldsymbol{g}=2\sqrt{\det\boldsymbol{g}}\sum_{k=1}^{n}\Omega_{ik}\cdot\frac{\partial x_{n+1}}{\partial x_{k}},\quad i=1,\dots,n.\label{eq: gradient metric determinant}
\end{equation}
one has the Christoffel symbols 
\begin{equation}
\Gamma_{ij}^{k}=\frac{\bar{f}_{k}\Omega_{ij}}{\sqrt{\det\boldsymbol{g}}},\quad i,j,k=1,\dots,n\label{eq: Christoffel symbols}
\end{equation}
and Riemann curvature tensor  

\begin{equation}
R_{iklm}=\Omega_{il}\Omega_{km}-\Omega_{kl}\Omega_{im},\quad i,k,l,m=1,\dots,n.\label{eq: Riemann curvature tensor}
\end{equation}
that is the classical Gauss equation. For Ricci tensor and scalar
curvature one gets 
\begin{equation}
R_{ij}=(\mbox{Tr}\boldsymbol{\Omega})\cdot\Omega_{ij}-\frac{{\displaystyle \sum_{k,l=1}^{n}\bar{f}_{k}\Omega_{kl} \bar{f}_{l}}}{\det\boldsymbol{g}}\Omega_{ij}-(\Omega^{2})_{ij}+\frac{1}{4\det\boldsymbol{g}^2}\frac{\partial\det\boldsymbol{g}}{\partial x_{i}}\cdot\frac{\partial\det\boldsymbol{g}}{\partial x_{j}},\label{eq: Ricci tensor}
\end{equation}

\begin{equation}
R=(\mbox{Tr}\boldsymbol{\Omega})^{2}-\mbox{Tr}[\boldsymbol{\Omega}^{2}]+2\frac{{\displaystyle \sum_{i,j,k=1}^{n}\bar{f}_{i}\cdot(\Omega_{ik}\Omega_{kj}-\Omega_{kk}\cdot\Omega_{ij})\bar{f}_{j}}}{\det\boldsymbol{g}}\label{eq: scalar curvature}
\end{equation}
where $\left(\boldsymbol{\Omega}\right)_{ij}=\Omega_{ij}$. 

Finally, the Gauss-Kronecker curvature of the statistical hypersurface
$V_{n}$ is given by  
\begin{equation}
{\displaystyle K\doteq\frac{\det\boldsymbol{\Omega}}{\det\boldsymbol{g}}=\frac{{\displaystyle \det\left|\bar{f}_{\{ik\}}-\bar{f}_{i}\cdot\bar{f}_{k}\right|}}{{\displaystyle \left(1+\sum_{l=1}^{n}\bar{f}_{l}^{2}\right)^{\frac{n+2}{2}}}}}.\label{eq: Gauss-Kronecker curvature}
\end{equation}

Entropy ${\displaystyle S=-\sum_{\alpha=1}^{m}w_{\alpha}\ln w_{\alpha}}$,
fundamental quantity in statistical physics, has also a simple geometrical
meaning. Namely, 
\begin{equation}
S=x_{n+1}-\bar{f}\label{eq: entropy}
\end{equation}
where ${\displaystyle \bar{f}=\sum_{\alpha=1}^{m}w_{\alpha}f_{\alpha}}$,
i.e. it is the deviation of the point on the hypersurface from the
mean value of functions $f_{\alpha}$, $\alpha=1,\dots,m$. 

Entropy is closely connected also with the scalar product $\overrightarrow{X}\cdot\overrightarrow{N}$
of the position vector $\overrightarrow{X}$ for the point on the
hypersurface and the corresponding normal vector $\overrightarrow{N}$.
Indeed taking into account (\ref{eq: position vector and normal vector})
one has 
\begin{equation}
\sqrt{\det\boldsymbol{g}}\,\overrightarrow{X}\cdot\overrightarrow{N}=x_{n+1}-\sum_{i=1}^{n}x_{i}\bar{f}_{i}.\label{eq: scalar product position normal vector}
\end{equation}
So 
\begin{equation}
S=\sqrt{\det\boldsymbol{g}}\,\overrightarrow{X}\cdot\overrightarrow{N}+\sum_{\alpha=1}^{m}w_{\alpha}\cdot\left(\sum_{i=1}^{n}x_{i}\frac{\partial f_{\alpha}}{\partial x_{i}}-f_{\alpha}\right).\label{eq: relation entropy and scalar product}
\end{equation}
In particular, if all functions $f_{\alpha}$ are homogeneous functions
of $x_{1},\dots,x_{n}$ of degree $d$ then ${\displaystyle S=\sqrt{\det\boldsymbol{g}}\,\overrightarrow{X}\cdot\overrightarrow{N}+(d-1)\bar{f}}$.
For $d=1$ one has $\displaystyle S=\sqrt{\det\boldsymbol{g}}\,\overrightarrow{X}\cdot\overrightarrow{N}.$

\section{\label{sec: Ideal statistical hypersurfaces } Ideal
hypersurfaces }

For ideal macroscopic systems the energy is a sum of energies of individual
particles or molecules which have their own energy spectrum \cite{LL1980}.
In general, such ideal situations are represented by functions
$f_{\alpha}$ which are decomposed into the sum of functions depending
on separate groups of variables, i.e. 
\begin{eqnarray}
f_{\alpha}(x_{1},\dots,x_{n})&=f_{\alpha_{1}}(x_{1},\dots,x_{n_{1}})+f_{\alpha_{2}}(x_{n_{1}+1},\dots,x_{n_{2}})+\nonumber \\&\dots+f_{\alpha_{l}}(x_{n_{l}+1},\dots,x_{n})\label{eq: additive energies}
\end{eqnarray}
with some functions $f_{\alpha_{p}}$, $\alpha_{p}=1,\dots,m_{p}$
and $p=1,\dots,l$. For instance, $f_{\alpha}(x_{1},x_{2},x_{3},x_{4})=f_{\alpha_{1}}(x_{1},x_{2})+f_{\alpha_{2}}(x_{3,}x_{4})$.
In this case also the general mapping (\ref{eq: basic statistical mapping})
is effectively decomposed and the corresponging hypersurface has several special features connected
with effective separation of groups of variables $\{x\}_{p}$. 

Here we will consider the simplest version of such ideal situation
with linear functions $f_{\alpha}$, i.e. 

\begin{equation}
f_{\alpha}(\boldsymbol{x})=\sum_{i=1}^{n}a_{\alpha i}x_{i}+b_{\alpha},\quad\alpha=1,\dots,m,\label{eq: linear model}
\end{equation}
where $a_{\alpha i}$ and $b_{\alpha}$ are constants. 
In this case 
\begin{equation}
\bar{f}_{i}=\sum_{\alpha=1}^{m}w_{\alpha}a_{\alpha i}\doteq\bar{a}_{i},\quad i=1,\dots,n\label{eq: derivative in linear case}
\end{equation}
and hence one has  
\begin{equation}
g_{ij}=\delta_{ij}+\bar{a}_{i}\cdot\bar{a}_{j},\quad i,j=1,\dots,n,\label{eq: metric in linear case}
\end{equation}
\begin{equation}
\Omega_{ij}=\frac{\overline{a_{ij}^{2}}-\bar{a}_{i}\bar{a}_{j}}{\sqrt{1+\sum_{l=1}^{n}\bar{a}_{l}^{2}}},\quad i,j=1,\dots,n\label{eq: second fundamental form in linear case}
\end{equation}
where 
\begin{equation}
\overline{a_{ij}^{2}}\doteq\sum_{\alpha=1}^{m}w_{\alpha}a_{\alpha i}a_{\alpha j},\quad i,j=1,\dots,n\label{eq: correlation for hessian in linear case}
\end{equation}
and 
\begin{eqnarray} R&={\displaystyle\frac{\left(\sum_{i=1}^{n}\overline{a_{ii}^{2}}-\sum_{i=1}^{n}\bar{a}_{i}^{2}
\right)^{2}-\sum_{i,k=1}^{n}(\overline{a_{ik}^{2}}-\bar{a}_{i}\bar{a}_{k})^{2}}{1+\sum_{i=1}^{n}\overline{a}_{i}^{2}}}\nonumber\\
&{\displaystyle +2\frac{\sum_{i,k,l=1}^{n}\overline{a_{il}^{2}}\cdot\overline{a_{lk}^{2}}
\bar{a}_{i}\bar{a}_{k}-\overline{a_{ll}^{2}}\cdot\overline{a_{ik}^{2}}
\bar{a}_{i}\bar{a}_{k}-\bar{a}_{i}^{2}\cdot\bar{a}_{l}\bar{a}_{k}\overline{a_{lk}^{2}}
}{\left(1+\sum_{i=1}^{n}\overline{a}_{i}^{2}\right)^{2}}}\nonumber\\
&\displaystyle{+2\frac{\sum_{i,k,l=1}^{n}\overline{a_{ll}^{2}}\bar{a}_{i}^{2}\cdot\bar{a}_{k}^{2}-\overline{a_{il}^{2}}\bar{a}_{i}\bar{a}_{l}
\cdot\bar{a}_{k}^{2}+\overline{a_{ik}^{2}}
\bar{a}_{i}\bar{a}_{k}\bar{a}_{l}^{2}}
{\left(1+\sum_{i=1}^{n}\overline{a}_{i}^{2}\right)^{2}}}
\label{eq: scalar curvature in linear case}
\end{eqnarray}
All these expressions contain the matrix $\left(\overline{a_{ij}^{2}}-\bar{a}_{i}\bar{a}_{j}\right)_{i,j}$ which can be rewritten as 
\begin{equation}
{\displaystyle \left(\overline{a_{ij}^{2}}-\bar{a}_{i}\bar{a}_{j}\right)_{i,j}=(\boldsymbol{a}^{T}H\boldsymbol{a})_{ij}}\label{eq: hessian linear case}
\end{equation}
where 
\begin{equation}
H_{\alpha\beta}\doteq\delta_{\alpha\beta}w_{\alpha}-w_{\alpha}w_{\beta},\quad\alpha,\beta=1,\dots,m.\label{eq: Hessian super-ideal}
\end{equation}
The Gauss-Kronecker curvature then is 
\begin{equation}
{\displaystyle K=\frac{{\displaystyle \det\left|\boldsymbol{a}^{T}H\boldsymbol{a}\right|}}{{\displaystyle \left(1+\sum_{l=1}^{n}\bar{a}_{l}^{2}\right)^{\frac{n+2}{2}}}}}.\label{eq: Gauss-Kronecker curvature in linear case}
\end{equation}
We note that if one considers a vector of $n$ random variables $\boldsymbol{X}=(X_{i})$,
$i=1,\dots,n$, which takes values $(a_{\alpha1},\dots,a_{\alpha n})$
with probability $w_{\alpha}$, then (\ref{eq: hessian linear case})
is the covariance matrix of $\boldsymbol{X}$. We also observe that
in this ideal case a hypersurface $V_{n}$ has non-trivial characteristics.
In general Riemann curvature tensor, scalar curvature and Gauss-Kronecker
curvature are different from zero. A sharp difference between this
result and those of Geometrothermodynamics \cite{QSTV2010,G-AMTdC2014}
is noted. 

Particular choices of the constants $a_{\alpha i}$ provide us with
special ideal hypersurfaces $V_{n}$. In particular, due to the presence
of the matrix $\boldsymbol{a}^{T}H\boldsymbol{a}$ in formulae (\ref{eq: second fundamental form in linear case})-(\ref{eq: Gauss-Kronecker curvature in linear case}), the rank of the matrix $\boldsymbol{a}$ (in general, rectangular
$m\times n$ matrix) plays fundamental role in characterization of
algebraic properties of Christoffel symbols, Riemann curvature tensor
and Gauss-Kronecker curvature. First, we observe that in virtue of
the normalization condition ${\displaystyle \sum_{\alpha=1}^{m}w_{\alpha}=1}$,
$\det H=0$. One also has  
\begin{prop}
\label{prop: Rank Hessian super-ideal case} The matrix $H$ is positive semidefinite and has rank $m-1$. \end{prop} 
\textit{Proof: }  
The Hessian matrix of a linear model is positive semidefinite since
it is a covariance matrix. In particular, $H$ is positive semidefinite since it is the Hessian matrix in the case ${\displaystyle F(\boldsymbol{x})=\log\left(\sum_{i=1}^{m}e^{x_{i}}\right)}$. 
Let $\boldsymbol{\zeta}=(\zeta_{1},\dots,\zeta_{m})^{T}\neq\boldsymbol{0}$
be an eigenvector of $H$: one has ${\displaystyle \sum_{j=1}^{m}(\delta_{ij}w_{i}-w_{i}w_{j})\zeta_{j}}$ ${\displaystyle=w_{i}\zeta_{i}-w_{i}\sum_{j=1}^{m}w_{j}\zeta_{j}}$
$\doteq w_{i}\zeta_{i}-w_{i}\overline{\zeta}$. If $\boldsymbol{\zeta}$
is a null eigenvector then ${\displaystyle w_{i}\zeta_{i}=w_{i}\overline{\zeta}}$.
Since $w_{i}\neq0$ by hypothesis, one gets $\zeta_{i}=\overline{\zeta}$
for all $i=1,\dots,m$. Then all $\zeta_{i}$ are equal, so $\boldsymbol{\zeta}=\overline{\zeta}\cdot(1,1,\dots,1)^{T}$
is the unique eigenvector for eigenvalue $0$ up to a constant $\overline{\zeta}$.
All other eigenvalues are strictly positive, so $\mbox{rank}H=m-1$.  \hfill\ensuremath{\square}

Then, due to standard properties of the matrices (see e.g. \cite{Gantmacher1977}),
one has 
\begin{equation}
\mbox{rank}(\boldsymbol{a}^{T}H\boldsymbol{a})\leq\min\{\mbox{rank}(\boldsymbol{a}),m-1\}.\label{eq: rank and products}
\end{equation}
Since $\mbox{rank}(\boldsymbol{a})\leq\min\{m,n\}$ one has 
\begin{equation}
\mbox{rank}(\boldsymbol{a}^{T}H\boldsymbol{a})\leq\min\{n,m-1\}.\label{eq: rank and products 2}
\end{equation}
So, if $n\geq m$, then $\mbox{rank}(\boldsymbol{a}^{T}H\boldsymbol{a})\leq m-1<n$
and, hence, the matrix $\boldsymbol{a}^{T}H\boldsymbol{a}$ is degenerate.
In particular, in this case the Gauss-Kronecker curvature vanishes  
\begin{equation}
K=0.\label{eq: K=0}
\end{equation}
and the second fundamental form is degenerate. 

If instead $m>n$, then Gauss-Kronecker
curvature for an ideal model (\ref{eq: linear model}) vanishes if
and only if there exists $\vec{x}_{0}=(x_{01},\dots,x_{0n})\neq(0,\dots,0)$
such that ${\displaystyle f_{\alpha}(\vec{x})-b_{\alpha}=\sum_{i=1}^{n}a_{\alpha i}x_{0i}}$
is independent of $\alpha$ for $\alpha=1,\dots,m$. Indeed, if $\mbox{rank}(\boldsymbol{a})<n$
then Gauss-Kronecker curvature is zero since $\mbox{rank}(\boldsymbol{a}^{T}H\boldsymbol{a})\leq\mbox{rank}(\boldsymbol{a})<n$.
In this case there exists a vector $\vec{x}_{0}=(x_{01},\dots,x_{0n})\neq(0,\dots,0)$
such that ${\displaystyle \sum_{i=1}^{n}a_{\alpha i}x_{0i}=0}$ for
all $\alpha=1,\dots,m$. At this point ${\displaystyle x_{n+1}\equiv\ln\left(\sum_{\alpha=1}^{m}e^{b_{\alpha}}\right)}$.
Then assume that $\mbox{rank}(\boldsymbol{a})=n$ and let us denote
$\vec{o}_{m}\doteq(\underset{m}{\underbrace{1,1,\dots,1}})^{T}$. If there exists $\vec{x}_{0}$ in $\mathbb{R}^{n}$ such that $\boldsymbol{a}\cdot\vec{x}_{0}=\vec{o}_{m}$,
i.e. ${\displaystyle \sum_{i=1}^{n}a_{\alpha i}x_{0i}=1}$ for $\alpha=1,\dots,m$,
then $\vec{x}_{0}$ is a null eigenvector for $\boldsymbol{a}^{T}H\boldsymbol{a}$
and $\det(\boldsymbol{a}^{T}H\boldsymbol{a})=0$. On the other hand,
suppose that $\boldsymbol{a}\cdot\vec{x}$ is not proportional to
$\vec{o}_{m}$ for all $\vec{x}$ in $\mathbb{R}^{n}$ and consider
$(\boldsymbol{a}\cdot\vec{x})^{T}\cdot H\cdot(\boldsymbol{a}\cdot\vec{x})=\vec{x}^{T}\cdot(\boldsymbol{a}^{T}H\boldsymbol{a})\cdot\vec{x}$
for a generic vector $\vec{x}$. We know from Proposition \ref{prop: Rank Hessian super-ideal case}
that this quantity is non-negative and it vanishes if and only if
$\boldsymbol{a}\cdot\vec{x}$ is proportional to $\vec{o}_{m}$. But this
contradicts our assumption, hence $\vec{x}^{T}\cdot(\boldsymbol{a}^{T}H\boldsymbol{a})\cdot\vec{x}$
is strictly positive for all $\vec{x}$ in $\mathbb{R}^{n}$. This
means that $\det(\boldsymbol{a}^{T}H\boldsymbol{a})>0$. 

For general ideal hypersurface $V_{n}$ with $f_{\alpha}$ given by
(\ref{eq: linear model}) and $b_{\alpha}=0$ for all $\alpha=1,\dots,m$
the entropy $S$ is given by formula (\ref{eq: entropy and scalar product}). 

There is one particular case of ideal statistical hypersurfaces closely
connected with the multi-soliton solutions of Korteweg\textendash de Vries (KdV) and  Kadomtsev\textendash Petviashvili
(KP) II equations. Indeed, with the
choice $\displaystyle a_{\alpha i}=\sum_{l=1}^N \eta_{\alpha l}(p_{l-1})^{2i-1}$, $i=1,2$, $\alpha=1,\dots,2^N$ where $p_{l}$ are arbitrary
constants and rows of the matrix $\eta_{\alpha l}$ represent all possible distributions of $0$ and $1$, one has $x_{3}=\log\tau$ where $\tau$ is the
tau-function of KdV equation \cite{Hirota1980}. 
In the generic case of all distinct $p_l$ $\mbox{rank}(\boldsymbol{a})=\min\{2,2^N\}=2$ and  $\mbox{rank}(\boldsymbol{a}^TH\boldsymbol{a})=\min\{2,2^N-1\}$. So for a statistical surface defined by one-soliton $\log \tau$ ($N=1$) the Gauss curvature vanishes. At the multi-soliton cases ($N\geq 2$) generically $K\neq 0$. For the multi-soliton solutions common for the $M$ first KdV flows ($i=1,2,\dots,M+1$) $\mbox{rank}(\boldsymbol{a})=\min \{M+1,2^N\}$. So, at sufficiently large $M$ $\mbox{rank}(\boldsymbol{a}^TH\boldsymbol{a})$ can be smaller than $M+1$ and, hence, for the corresponding statistical hypersurface the Gauss-Kronecker curvature vanishes. 

Multi-soliton $\log \tau$ for the  Kadomtsev\textendash Petviashvili (KP) II equation and hierarchy also correspond to this case with more complicated form of rectangular matrices $a_{\alpha i}$ (see e.g. \cite{BC2006,CK2008}). Again one-soliton statistical hypersurface has Gauss-Kronecker curvature equal to zero while there are various different cases for general $(N,M)$ solitons. This problem will be discussed in details elsewhere. 

Finally, we note that in the case when all $a_{\alpha i}$ are positive integers $n_{\alpha i}$ the ideal mapping (\ref{eq: basic statistical mapping})-(\ref{eq: linear model}) in terms of variables $y_{i}$ defined as $x_i=\log y_i$, $i=1,\dots, n+1$, turns to a pure algebraic one 
\begin{equation}
y_{n+1}=\sum_{\alpha=1}^m c_{\alpha} \prod_{i=1}^n (y_i)^{n_{\alpha i}}
\label{eq: algebraic form and occupation numbers}
\end{equation}
where $c_{\alpha}\doteq e^{b_{\alpha}}$. In physical context such $n_{\alpha i}$ have clear meaning of occupation numbers and $\bar{a}_{i}=\overline{n_{i}}$ are their mean values. In geometry the formula (\ref{eq: algebraic form and occupation numbers}) provides us with algebraic hypersurfaces (if one treats $y_1,\dots,y_{n+1}$ as local coordinates in $\mathbb{R}^{n+1}$). 

\section{\label{sec: Special ideal hypersurfaces} Super-ideal hypersurfaces}

In the special ideal case with $m=n$ and $a_{\alpha i}=\delta_{\alpha i}$,
$b_{\alpha}=0$ where $\delta_{\alpha i}$ is the Kronecker symbol
all formulae are drastically simplified. Indeed, one has 
\begin{equation}
{\displaystyle x_{n+1}=\ln\left(\sum_{i=1}^{n}e^{x_{i}}\right)}\label{eq: super ideal hypergraph}
\end{equation}
and hence  
\begin{equation}
g_{ij}=\delta_{ij}+w_{i}w_{j},\quad i,j=1,2,\dots,n,\label{eq: special ideal mapping metric}
\end{equation}

\begin{equation}
\Omega_{ij}=\frac{w_{i}(\delta_{ij}-w_{j})}{\sqrt{1+S_{2}}},\label{eq: special ideal mapping second fundamental form}
\end{equation}

\begin{eqnarray}
R_{sijk}&=\frac{1}{1+S_{2}}\left(\delta_{js}\delta_{ik}w_{i}w_{j}-\delta_{js}w_{i}w_{j}w_{k}-\delta_{ks}\delta_{ij}w_{i}w_{k}
+\delta_{ks}w_{i}w_{j}w_{k}\right.\nonumber \\ &\left.+\delta_{ij}w_{i}w_{k}w_{s}-\delta_{ik}w_{i}w_{j}w_{s}\right),\label{eq: special ideal mapping covariant Riemann tensor}
\end{eqnarray}

\begin{eqnarray}
R_{ik}&=-\frac{1}{(1+S_{2})^2}\left[((1+S_{2})(1+\delta_{ik})-(1+w_{i})(1+w_{k}))w_{i}w_{k}\right.\nonumber\\
&\left.+(S_{3}-1)(\delta_{ik}w_{i}-w_{i}w_{k})\right],\label{eq: special ideal mapping Ricci tensor}
\end{eqnarray}

\begin{equation}
R=\frac{2(1+S_{4})}{(1+S_{2})^2}-1\label{eq: special ideal mapping scalar curvature}
\end{equation}
and mean curvature (see e.g. \cite{Eisenhart1997}) 

\begin{equation}
\Omega\doteq\sum_{i,j=1}^n g^{ij}\Omega_{ij}=\sum_{i,j=1}^n \left(\delta_{ij}-\frac{w_{i}w_{j}}{1+S_{2}}\right)\Omega_{ij}=\frac{1-S_{3}}{(1+S_{2})^{\frac{3}{2}}}\label{eq: special ideal mapping mean curvature}
\end{equation}
where $S_{p}$ are power sums   
\begin{equation}
S_{p}\doteq\sum_{i=1}^{n}w_{i}^{p}.\label{eq: symmetric sums}
\end{equation}
Finally, Gauss-Kronecker curvature is  
\begin{equation}
K=0.\label{eq: special ideal mapping Gauss-Kronecker curvature}
\end{equation}
Vanishing of the Gauss-Kronecker curvature for the super-ideal case
is connected with the fact that hypersurface $V_{n}$ defined by (\ref{eq: special ideal mapping metric})
admits the symmetry transformation 
\begin{equation}
x_{i}\mapsto x'_{i}=x_{i}+a,\quad i=1,\dots,n+1\label{eq: global translation}
\end{equation}
with arbitrary parameter $a$ and corresponding Killing vector is ${\displaystyle \mathcal{K}=\sum_{i=1}^{n}\frac{\partial}{\partial x_{i}}}$. 
We note that the probabilities $w_{i}$ are invariant under the transformation
(\ref{eq: global translation}). 

All geometric characteristics of super-ideal hypersurface are algebraic
functions of the Gibbs probabilities ${\displaystyle w_{i}=\frac{e^{x_{i}}}{\sum_{j=1}^{n}e^{x_{j}}}}$.
These variables $\vec{w}=(w_{1},\dots,w_{n})$ obey the constraint
${\displaystyle S_{1}=\sum_{i=1}^{n}w_{i}=1}$ and vary in the intervals
which depend on intervals of variations of variables $x_{1},\dots,x_{n}$.
Here we will consider the case when all $x_{i}$ are unbounded. So $0\leq w_{i}\leq1$, $i=1,\dots,n$ and the point
$\vec{w}$ belong to the hyperplane ${\displaystyle \sum_{i=1}^{n}w_{i}=1}$
passing through the vertices $\vec{e}_{\alpha}$ ($(\vec{e}_{\alpha})_{i}=\delta_{\alpha i}$,
$\alpha,i=1,\dots,n$) of the unit $n$-dimensional cube in $\mathbb{R}^{n}$. 

Scalar and mean curvatures (\ref{eq: special ideal mapping scalar curvature})-(\ref{eq: special ideal mapping mean curvature}) have special properties due
to their simple dependence only on the power sums $S_{2}$, $S_{3}$,
$S_{4}$. Indeed, one has 
\begin{prop}
\label{prop: Scalar curvature fundamental case positive}Mean and scalar curvatures of super-ideal statistical hypersurfaces
take values in the intervals 
\begin{equation}
0\leq\Omega\leq\frac{n-1}{\sqrt{n(n+1)}},\quad\quad 0\leq R\leq\frac{(n-1)(n-2)}{n(n+1)}. \label{eq: bounds mean and scalar curvature}
\end{equation}\end{prop}
\textit{Proof: }
First we observe that the classical power mean (Hölder) inequality
(see e.g. \cite{HardyLittlewoodPolya1988}) 
\begin{equation}
\left(\frac{\sum_{i=1}^{n}w_{i}^{p}}{n}\right)^{\frac{1}{p}}\geq\frac{\sum_{i=1}^{n}w_{i}}{n}\label{eq: means inequality}
\end{equation}
with integer $p\geq1$ in our case ($S_{1}=1$) implies 
\begin{equation}
S_{p}\geq\frac{1}{n^{p-1}}.\label{eq: bound power sum}
\end{equation}
Hence, the power sums are bounded 
\begin{equation}
\frac{1}{n^{p-1}}\leq S_{p}\leq1,\quad p=2,3,4\dots\label{eq: bounds power sum}
\end{equation}

The maximum value $(S_{p})_{\max}=1$ is achieved at the vertex poins
$\vec{e}_{\alpha}$, $\alpha=1,\dots,n$ while ${\displaystyle (S_{p})_{\min}=\frac{1}{n^{p-1}}}$
at the point ${\displaystyle \vec{e}_{0}=\left(\frac{1}{n},\dots,\frac{1}{n}\right)}$. 

The values of $\Omega$ and $R$ at these particular points provide us 
also with their lower and upper bounds. Since $S_{3}\leq1$ one immediately
concludes from the formula (\ref{eq: special ideal mapping mean curvature}) that the mean curvature $\Omega\geq0$
and $\Omega|_{\vec{e}_{\alpha}}=0$, $\alpha=1,\dots,n$. 

Then for
the maximum of $\Omega$ one gets 
\begin{equation} \Omega_{\max}=\frac{\max\{1-S_{3}\}}{\min\{(1+S_{2})^{\frac{3}{2}}\}}=\frac{1-\min\{S_{3}\}}{(1+\min\{S_{2}\})^{\frac{3}{2}}}
=\frac{n-1}{\sqrt{n(n+1)}}=\Omega|_{\vec{e}_{0}}.
\label{eq: max mean curvature super-ideal case}
\end{equation} 

For the scalar curvature one also has $R|_{\vec{e}_{\alpha}}=0$,
$\alpha=1,\dots,n$. In order to prove that $R\geq0$ it is sufficient
to show that 
\begin{equation}
\hat{R}(w_{1},\dots,w_{n})\doteq{\displaystyle 2S_{4}+2-(1+S_{2})^{2}\geq0}\label{eq: numerator scalar curvature}
\end{equation}
First, one can show that for any $2\leq i\leq n$ 
\begin{eqnarray}
\hat{R}(w_{1},\dots,w_{n})&-\hat{R}(w_{1}+w_{i},w_{2},\dots,w_{i-1},0,w_{i+1},\dots,w_{n})\nonumber\\
&=4w_{1}w_{i}\left[1-(w_{1}+w_{i})^{2}+\sum_{j\neq1,i}w_{j}^{2}\right]\geq0
\label{eq: inequality lemma}
\end{eqnarray}
since $0\leq w_{1}+w_{i}\leq1$. So one has a chain of inequalities
\begin{eqnarray}
\hat{R}(\vec{w})&\geq\hat{R}(w_{1}+w_{n},w_{2},\dots,w_{n-1},0)\nonumber\\
&\geq\hat{R}(w_{1}+w_{n}+w_{n-1},w_{2},\dots,w_{n-2},0,0)\geq\dots\nonumber\\
&\geq\hat{R}(w_{1}+w_{2}+\dots+w_{n},0,0,\dots,0)=\hat{R}(1,0,\dots,0).
\label{eq: chain inequalities}
\end{eqnarray}
Since $\hat{R}(1,0,\dots,0)=0$ one gets (\ref{eq: numerator scalar curvature})
and, consequently, $R\geq0$. 

One can also show that 
\begin{equation}
\hat{R}(\vec{w}_{(ij)})-\hat{R}(\vec{w})=(w_{i}-w_{j})^{2}\cdot\left[1-(w_{i}+w_{j})^{2}+\sum_{k\neq i,j}w_{k}^{2}\right]\geq0\label{eq: lemma inequality 2}
\end{equation}
for any $i,j=1,\dots,n$ where 
\begin{equation} 
\vec{w}_{(ij)}\doteq\left(w_{1},\dots,w_{i-1},\frac{w_{i}+w_{j}}{2},w_{i+1},\dots,w_{j-1},\frac{w_{i}+w_{j}}{2},w_{j+1},
\dots,w_{n}\right).\label{eq: inequality scalar curvature upper bound}
\end{equation}
Then, since ${\displaystyle S_{2}(\vec{w})-S_{2}(\vec{w}_{(ij)})=\frac{(w_{i}-w_{j})^{2}}{2}\geq0}$
one has ${\displaystyle \frac{1}{\left(1+S_{2}(\vec{w}_{(ij)})\right)^{2}}\geq\frac{1}{(1+S_{2}(\vec{w}))^{2}}}$.
Hence 
\begin{equation}
R(\vec{w}_{(ij)})\geq R(\vec{w})\label{eq: inequality scalar curvature}
\end{equation}
for any $i,j=1,\dots,n$. 

The inequality (\ref{eq: inequality scalar curvature}) implies that
the maximum of $R$ is reached if $R(\vec{w}_{(ij)})=R(\vec{w})$ for all $i,j=1,\dots,n$. This happens at $w_{1}=w_{2}=\dots=w_{n}$, i.e.
at the point ${\displaystyle \vec{e}_{0}=\left(\frac{1}{n},\dots,\frac{1}{n}\right)}$.
Thus, $R_{\max}=R|_{\vec{e}_{0}}$, i.e. 
\begin{equation}
{\displaystyle R_{\max}=\frac{{\displaystyle 2\left(1+\frac{1}{n^{3}}\right)}}{{\displaystyle \left(1+\frac{1}{n}\right)^{2}}}=\frac{{\displaystyle (n-1)(n-2)}}{n{\displaystyle \left(n+1\right)}}}.\label{eq: max scalar curvature}
\end{equation} \ensuremath{\square}

Note also that 
\begin{equation}
\left(\sqrt{\det\boldsymbol{g}}\cdot\Omega\right)|_{\vec{e}_{0}}=\frac{n-1}{n}\label{eq: det g^(1/2)*mean at e_0}
\end{equation}
and 
\begin{equation}
\left(\det\boldsymbol{g}\cdot R\right)|_{\vec{e}_{0}}=\frac{(n-1)(n-2)}{n^{2}}.\label{eq: det g*scalar at e_0}
\end{equation}
The point $\vec{e}_{0}$ corresponds to the straight line $x_{1}=x_{2}=\dots=x_{n}$. 

Finally, in the super-ideal case (\ref{eq: special ideal mapping metric})
the normal vector is $\displaystyle \overrightarrow{N}=\frac{1}{\sqrt{1+S_{2}}}(-w_{1},\dots,-w_{n},1)$
and hence for entropy $S$ one has 
\begin{equation}
S(\boldsymbol{x})=x_{n+1}-\sum_{i=1}^{n}w_{i}x_{i}=\sqrt{1+S_{2}}\,\overrightarrow{X}\cdot\overrightarrow{N}.\label{eq: entropy special normal case}
\end{equation}
 So the normal vector has pure probabilistic character and the entropy
is the difference between $x_{n+1}$ and mean value $\bar{x}$.

\section{\label{sec: Non-ideal case} Non-ideal case}

Hypersurfaces with nonlinear and nonseparable functions $f_{\alpha}(\boldsymbol{x})$
correspond to macroscopic systems with interaction between particles,
molecula \textit{etc}. Properties of such non-ideal hypersurfaces
vary according to properties of functions $f_{\alpha}(\boldsymbol{x})$.
Here we consider few illustrative examples. 

The first case is $n=2$, $m=1$ and 
\begin{equation}
F(x_{1},x_{2})=x_{1}+x_{2}+\varphi(x_{1},x_{2})\label{eq: interacting, example 1}
\end{equation}
where $\varphi(x_{1},x_{2})$ is a function. 

One has a surface in $\mathbb{R}^{3}$ given by 
\begin{equation}
x_{3}=x_{1}+x_{2}+\varphi(x_{1},x_{2})\label{eq: interacting, example 1, mapping}
\end{equation}
 with all standard formulae for a surface. 

The second example corresponds to $n=3$, $m=2$ and 
\begin{equation}
x_{4}=\ln\left(e^{x_{1}+x_{2}+\varepsilon x_{1}x_{2}}+e^{x_{1}+x_{3}+\varepsilon x_{1}x_{3}}\right)\label{eq: interacting, example 2, mapping}
\end{equation}
where $\varepsilon$ is a constant. One has ${\displaystyle \frac{\partial x_{4}}{\partial x_{1}}=1+\varepsilon(w_{1}x_{2}+w_{2}x_{3})}$,
${\displaystyle \frac{\partial x_{4}}{\partial x_{2}}=w_{1}(1+\varepsilon x_{1})}$ 
and \\${\displaystyle \frac{\partial x_{4}}{\partial x_{3}}=w_{2}(1+\varepsilon x_{1})}$.
Hence, 
\begin{eqnarray}
g_{11}=1+[1+\varepsilon(w_{1}x_{2}+w_{2}x_{3})]^{2}, & g_{23}=w_{1}w_{2}(1+\varepsilon x_{1})^{2},\nonumber\\
g_{12}=(w_{1}+\varepsilon(w_{1}^{2}x_{2}+w_{1}w_{2}x_{3}))\cdot(1+\varepsilon x_{1}), & g_{22}=1+w_{1}^{2}(1+\varepsilon x_{1})^{2},\nonumber\\
g_{13}=(w_{2}+\varepsilon(w_{1}w_{2}x_{2}+w_{2}^{2}x_{3}))\cdot(1+\varepsilon x_{1}), & g_{33}=1+w_{2}^{2}(1+\varepsilon x_{1})^{2}
\label{eq: interacting, example 2, metric}
\end{eqnarray}
and 
\begin{eqnarray}
\det\boldsymbol{g}&=1+(1+\varepsilon(w_{1}x_{2}+w_{2}x_{3}))^{2}+(w_{1}^{2}+w_{2}^{2})(1+\varepsilon x_{1})^{2}\nonumber\\
&=\left[2+w_{1}^{2}+w_{2}^{2}\right]+\left[2(w_{1}x_{2}+w_{2}x_{3})+2(w_{1}^{2}+w_{2}^{2})\cdot x_{1}\right]\varepsilon\\
&+\left[(w_{1}^{2}+w_{2}^{2})\cdot x_{1}^{2}+(w_{1}x_{2}+w_{2}x_{3})^{2}\right]\varepsilon^{2}\nonumber
\label{eq: interacting, example 2, metric determinant}
\end{eqnarray}
where ${\displaystyle w_{1}=\frac{e^{x_{1}+x_{2}+\varepsilon x_{1}x_{2}}}{e^{x_{1}+x_{2}+\varepsilon x_{1}x_{2}}+e^{x_{1}+x_{3}+\varepsilon x_{1}x_{3}}}}$
and ${\displaystyle w_{2}=\frac{e^{x_{1}+x_{3}+\varepsilon x_{1}x_{3}}}{e^{x_{1}+x_{2}+\varepsilon x_{1}x_{2}}+e^{x_{1}+x_{3}+\varepsilon x_{1}x_{3}}}}$.
The Gauss-Kronecker curvature is given by 

\begin{equation}
{\displaystyle {\displaystyle K=-\frac{w_{1}w_{2}\cdot\varepsilon^{2}(1+\varepsilon x_{1})^{2}}{\det\boldsymbol{g}^{\frac{5}{2}}}}}.\label{eq: interacting, example 2, GK curvature}
\end{equation}

So the Gauss-Kronecker curvature is different from zero. It is connected
also with the fact that the formula (\ref{eq: interacting, example 2, mapping})
is not invariant under the shift $x_{i}\mapsto x_{i}+a$, $i=1,2,3,4$. Finally, for entropy one finds 

\begin{equation}
S=\ln(e^{x_{2}+\varepsilon x_{1}x_{2}}+e^{x_{3}+\varepsilon x_{1}x_{3}})-\frac{(1+\varepsilon x_{1})\cdot \left(e^{x_{2}+\varepsilon x_{1}x_{2}}\cdot x_{2}+e^{x_{3}+\varepsilon x_{1}x_{3}}\cdot x_{3}\right)}{e^{x_{2}+\varepsilon x_{1}x_{2}}+e^{x_{3}+\varepsilon x_{1}x_{3}}}.
\label{eq: interacting, example 2, entropy}
\end{equation}

It should be noted that Gauss-Kronecker curvature can be zero even
in non-ideal case. It happens if a hypersurface admits a translational
symmetry. For instance, if instead of (\ref{eq: interacting, example 2, mapping})
the hypersurface is defined by 
\begin{equation}
x_{4}=\ln\left(e^{x_{1}+x_{2}+\varepsilon(x_{1}-x_{2})^{2}}+e^{x_{1}+x_{3}+\varepsilon(x_{1}-x_{3})^{2}}\right)\label{eq: interacting, example 3, mapping}
\end{equation}
then it is invariant under the transformation $x_{i}\mapsto x'_{i}=x_{i}+a$,
$i=1,2,3$, $x_{4}\mapsto x'_{4}=x_{4}+2a$ and the corresponding
Gauss-Kronecker curvature is $K=0$. 

In physics it is often quite useful to
study first corrections to ideality for ``small'' interactions. For hypersurfaces it corresponds
to ``small'' nonlinearities. Let us consider the hypersurface given
by (\ref{eq: interacting, example 2, mapping}) with $0<\varepsilon\ll1$
and calculate first order corrections in $\varepsilon$ to ideal case.
Denoting the first order corrections in $\varepsilon$ of the function
$f$ by $I_{\varepsilon}[f]$, we find 

\begin{eqnarray}
I_{\varepsilon}[g_{11}]=2(w_{1}x_{2}+w_{2}x_{3}), & I_{\varepsilon}[g_{23}]=2w_{1}w_{2}x_{1},\nonumber\\
I_{\varepsilon}[g_{12}]=w_{1}^{2}x_{2}+w_{1}w_{2}x_{3}+w_{1}x_{1}, & I_{\varepsilon}[g_{22}]=2w_{1}^{2}x_{1},\\
I_{\varepsilon}[g_{13}]=w_{1}w_{2}x_{2}+w_{2}^{2}x_{3}+w_{2}x_{1}, & I_{\varepsilon}[g_{33}]=2w_{2}^{2}x_{1}\nonumber
\label{eq: interacting, example 2, metric, first order correction}
\end{eqnarray}
and 
\begin{equation}
I_{\varepsilon}[\det\boldsymbol{g}]=2(w_{1}x_{2}+w_{2}x_{3})+2(w_{1}^{2}+w_{2}^{2})\cdot x_{1},\label{eq: interacting, example 2, metric determinant, first order correction}
\end{equation}

\begin{equation}
{\displaystyle {\displaystyle I_{\varepsilon}[K]=0}.}\label{eq: interacting, example 2, GK curvature, first order correction}
\end{equation}

Second order correction for Gauss-Kronecker curvature does not vanishes. It can be seen from (\ref{eq: interacting, example 2, GK curvature})
and (\ref{eq: interacting, example 2, metric determinant}) that second
order series coefficient is 
\begin{equation}
{\displaystyle II_{\varepsilon}[K]=-\frac{e^{x_{2}+x_{3}}(e^{x_{2}}+e^{x_{3}})^{3}}{(3e^{2x_{2}}+4e^{x_{2}+x_{3}}+3e^{2x_{3}})^{\frac{5}{2}}}}.\label{eq: interacting, example 2, GK curvature, second order correction}
\end{equation}

Next let us consider the problem of the vanishing of first order corrections in a
slighty more general case with a physical perspective. So, let $P$
be a number of different subsystems $\mathcal{S}_{p}$, $p=1,\dots,P$ 
with $q_{p}$ different levels in subsystem $\mathcal{S}_{p}$ labeled
by $(x_{i}^{p})$, where $x_{i}^{p}$ is the $i$-th level in the
$p$-th subsystems, $i=1,2,\dots,q_{p}$. Let us call the total number
of levels $n={\displaystyle \sum_{p=1}^{P}q_{p}}$ and put $\vec{x}_{p}\doteq(x_{1}^{p},x_{2}^{p},\dots,x_{q_{p}}^{p})$,
$p=1,\dots,P$. 

These systems interact via a function ${\displaystyle \Gamma(x_{i_{1}}^{1},x_{i_{2}}^{2},\dots,x_{i_{P}}^{P})}$
of $P$ variables. Thus the statistical mapping is given by 

\begin{equation}
{\displaystyle F(\boldsymbol{x})=\ln\left(\sum_{s:\,\mathcal{P}\longrightarrow Q}e^{\Gamma(x_{s(1)}^{1},x_{s(2)}^{2},\dots,x_{s(P)}^{P})}\right),}\label{eq: basic statistical mapping with interactions}
\end{equation}
where the sum is over all possible mappings $s:\,\{1,2,\dots P\}\longrightarrow{\displaystyle \bigcup_{p=1}^{P}\mathcal{S}_{p}}$
such that $s(p)$ belongs to subsystem $\mathcal{S}_{p}$ for all
$p=1,\dots,P$. Functions $f_{\alpha}$ are now parametrized by the
set $\mathcal{C}$ of all these mappings $s$. So, $m=\#\mathcal{C}=q_{1}\cdot q_{2}\cdot\dots\cdot q_{P}$
and Gibbs weight corresponding to $s$ is 
\begin{equation}
{\displaystyle w_{s}\doteq\frac{e^{\Gamma(x_{s(1)}^{1},x_{s(2)}^{2},\dots,x_{s(P)}^{P})}}{{\displaystyle \sum_{\mbox{ \ensuremath{t}\ensuremath{\in\mathcal{C}}}}e^{\Gamma(x_{t(1)}^{1},x_{t(2)}^{2},\dots,x_{t(P)}^{P})}}}.}\label{eq: basic statistical mapping with interactions, probabilities}
\end{equation}

In order to study deviation from ideality, we have to define what
an ideal linear model of $P$ non-interacting subsystems is. Due to
the relations (\ref{eq: additive energies})
and (\ref{eq: linear model}), it is natural to consider possible
energies as $i(x_{1},x_{2},\dots,x_{P})={\displaystyle \sum_{p=1}^{P}}c_{p}x_{p}$
for real constants $c_{p}$ and all possible choices of $x_{p}$ in
the $p$-th subsystem. So ideal statistical mapping is ${\displaystyle F_{0}(\vec{x}_{1},\dots,\vec{x}_{P})=\ln\left({\displaystyle \sum_{t\in\mathcal{C}}\prod_{p=1}^{P}e^{i(x_{t(t)}^{1},\dots,x_{t(P)}^{P})}}\right)}$
${\displaystyle =\ln\left[\prod_{p=1}^{P}\left(\sum_{i=1}^{q_{p}}e^{c_{p}x_{i}^{p}}\right)\right]}$
that is 
\begin{equation}
F_{0}(\vec{x}_{1},\vec{x}_{2},\dots,\vec{x}_{P})=\sum_{p=1}^{P}\varphi_{p}(\vec{x}_{p})\label{eq: ideal p-subsystems mapping}
\end{equation}
where 
\begin{equation}
\varphi_{p}(\vec{x}_{p})\doteq\ln\left(\sum_{i=1}^{q_{p}}e^{c_{p}x_{i}^{p}}\right),\quad p=1,\dots,P.\label{eq: partial statistical functions}
\end{equation}
Then Hessian matrix is block diagonal and the model is ideal since
it is linear and Gauss-Kronecker curvature vanishes: indeed, vector
$(\underset{n}{\underbrace{1,1,\dots,1}})^{T}$ is an eigenvector
of Hessian matrix with eigenvalue $0$. 

Perturbation of this model means passing from $i(x_{1},x_{2},\dots,x_{P})$
to $i(x_{1},x_{2},\dots,x_{P})+\varepsilon\cdot\gamma(x_{1},x_{2},\dots,x_{P})$.
So one has 
\begin{prop}
\label{prop: First order correction curvature} For ``small'' perturbations
of the form $\varepsilon\cdot\gamma(x_{1},x_{2},\dots,x_{P})$, where
$\gamma(x_{1},x_{2},\dots,x_{P})$ is any smooth function and $0<\varepsilon\ll1$,
the first order correction in $\varepsilon$ to the Gauss-Kronecker curvature
of the statistical hypersurface defined by ${\displaystyle F(\vec{x}_{1},\dots,\vec{x}_{P})=\ln\left[\sum_{t\in\mathcal{C}}\exp(j(x_{t(1)}^{1},x_{t(2)}^{2},\dots,x_{t(P)}^{P}))\right]}$
with $j(x_{1},x_{2},\dots,x_{P})={\displaystyle \sum_{p=1}^{P}}c_{p} x_{p}+\varepsilon\cdot\gamma(x_{1},x_{2},\dots,x_{P})$
vanishes. \end{prop}
\textit{Proof: }
See Appendix \ref{sec: Appendix A }. 
\hfill\ensuremath{\square}

\section{\label{sec: Singularities of the statistical hypersurface } Singularities
of the statistical hypersurface }

Connection of hypersurfaces $V_{n}$ with statistical physics suggests
to analyze non-smooth behaviour analogous to that typical for phase
transitions. 

Following the standard classification of phase transitions (see e.g.
\cite{LL1980}) we will refer to a singularity of hypersurface
$V_{n}$ for which all derivatives of $F$ of order $0,1,\dots,k-1$
are continuous and at least one derivative of order $k$ is discontinuous
as $k$-th order phase singularity. For the singularity of the first
order a hypersurface is smooth while the metric and second form are discontinuous. For example, hypersurface has an edge
and metric and second form exhibit a jump along this edge. At the
same time curvature, in general, blows up. For the second order singularity
hypersurface, metric and normal vector are smooth while curvature
has a jump. Due to a rather complicated expression for the Riemann
curvature tensor and Gauss-Kronecker curvature they may have no blow
up. We will refer to such singularities as hidden. 

Ideal hypersurfaces defined by (\ref{eq: linear model}) clearly do
not have such phase singularities. On the other hand an analysis of
case of general nonlinear functions $f_{\alpha}(\boldsymbol{x})$
is rather involved. Here we will discuss few examples of non-ideal
statistical hypersurface in order to illustrate some properties of
hidden and visible singularities. First order phase singularities
are considered in first two examples.

\paragraph{\label{exa: Visible first order} First example:}
visible singularities at $n=2$. Let $x_{1}=x$ and $x_{2}=y$
be coordinates and $\{f_{\alpha}(x,y)\}$ for $\alpha=2,\dots,m$
be smooth functions, e.g. linear functions ${\displaystyle f_{\alpha}(x,y)=c_{\alpha1}x+c_{\alpha2}y}$.
Then let us consider function ${\displaystyle f_{1}(x,y)\doteq s(x)+h(y)}$
with 
\begin{equation}
{\displaystyle s(x)=\sqrt[3]{(x-x_{0})^{4}}+(x-x_{0})\cdot\Theta(x-x_{0})},\label{eq: Visible first order phase singularity, diverging Gauss-Kronecker and scalar curvature}
\end{equation}
where $\Theta(x)$ is Heaviside step function and $h(y)$ is a smooth
function such that there exist points $\{\bar{y}\}$ where $h''(\bar{y})>0$.
Here we denote ${\displaystyle h'(\bar{y})=c_{12}}$. Then
${\displaystyle \frac{\partial f_{1}}{\partial y}}$, ${\displaystyle \frac{\partial f_{\alpha}}{\partial x}}$
and ${\displaystyle \frac{\partial f_{\alpha}}{\partial y}}$ are
continuous for all $\alpha\geq2$ and ${\displaystyle \left(\frac{\partial f_{1}}{\partial x}\right)^{2}}$
is finite and discontinuous at $x=x_{0}$. So $g_{11}$ and $\det\boldsymbol{g}$
have a jump here, i.e. it is a first order phase singularity. Even if $F$ is not differentiable at $x=x_{0}$ one can study the behaviour of its Hessian determinant in a neighborhood of this point. First, ${\displaystyle \frac{\partial^{2}F}{\partial x\partial y}}$
is finite since all second derivatives ${\displaystyle \frac{\partial^{2}f_{\alpha}}{\partial x\partial y}=0}$
for $\alpha=1,\dots,m$ and first derivatives are finite. Then, ${\displaystyle \frac{\partial^{2}F}{\partial y^{2}}=w_{1}h''+\sum_{\alpha=1}^{m}w_{\alpha}c_{\alpha2}^{2}-\left(\sum_{\alpha=1}^{m}w_{\alpha}c_{\alpha2}\right)^{2}}$ and
at all points in $\{\bar{y}\}$ we have $h''(\bar{y})>0$, so \\
 ${\displaystyle \left.\frac{\partial^{2}F}{\partial y^{2}}\right|_{y=\bar{y}}>\sum_{\alpha=1}^{m}w_{\alpha}c_{\alpha2}^{2}-\left(\sum_{\beta=1}^{m}w_{\beta}c_{\beta2}\right)^{2}=\sum_{\alpha=1}^{m}w_{\alpha}\left(c_{\alpha2}-\sum_{\beta=1}^{m}w_{\beta}c_{\beta2}\right)^{2}\geq0}$.
Finally, \\
${\displaystyle \frac{\partial^{2}F}{\partial x^{2}}=w_{1}s''(x)+w_{1}(s'(x))^{2}+\sum_{\alpha=2}^{m}w_{\alpha}(c_{\alpha1})^{2}-\left(\bar{f}_{1}\right)^{2}}$.
Last three terms are everywhere finite while ${\displaystyle \lim_{x\rightarrow x_{0}+}s''(x)=\lim_{x\rightarrow x_{0}-}s''(x)=+\infty}$.
Hence, Hessian determinant ${\displaystyle \frac{\partial^{2}F}{\partial x^{2}}\frac{\partial^{2}F}{\partial y^{2}}-\left(\frac{\partial^{2}F}{\partial x\partial y}\right)^{2}}$
is equal to ${\displaystyle w_{1}\frac{\partial^{2}F}{\partial y^{2}}s''(x)}$
plus some terms which stay finite for $x\rightarrow x_{0}$. So, ${\displaystyle \lim_{x\rightarrow x_{0}}w_{1}\frac{\partial^{2}F}{\partial y^{2}}s''(x)=+\infty}$
at all point $\{\bar{y}\}$ and Gauss curvature $K$ and scalar
curvature $2K$ diverge at $x\rightarrow x_{0}$.

\paragraph{\label{exa: Hidden first order} Second example: }

hidden first order phase singularity. Let $A(x)$ be a $m\times(n-1)$
matrix whose entries are smooth functions of $x$ of the type ${\displaystyle A_{\alpha i}(x)=c_{\alpha i}+p_{\alpha i}(x)}$,
$c_{\alpha i}$ are real constants and all $p_{\alpha i}(x)$ tend to
$0$ as $x\rightarrow x_{0}$. More precisely, let the matrix ${\displaystyle A_{0}\doteq\left(c_{\alpha i}\right)}_{\alpha=1,\dots,m}^{i=2,\dots,n}$
corresponds to an ideal model with $\det A_{0}=0$ and $p_{\alpha i}(x)\sim q_{\alpha i}\cdot(x-x_{0})^{2}$
for $x\sim x_{0}$, with real constants $q_{\alpha i}$, $\alpha=1,\dots,m$
and $i=2,\dots,n$. Then, we introduce the function  
\begin{eqnarray}
s(x)&{\displaystyle =\left(x-\frac{x}{2}\cdot\sqrt{1-\frac{x^{2}}{x_{0}^{2}}}-\frac{x_{0}\cdot\arcsin(\frac{x}{x_{0}})}{2}\right)
\Theta(x_{0}^{2}-x^{2})}\nonumber\\
&{\displaystyle +x_{0}\frac{4-\pi}{4}\Theta(x-x_{0})-x_{0}\frac{4-\pi}{4}\Theta(-x_{0}-x)}.\label{eq: check function}
\end{eqnarray}
The statistical model is given by the $y$-linear system ${\displaystyle \tilde{f}_{\alpha}(x,y_{2},\dots,y_{n})=\sum_{i=2}^{n}A_{\alpha i}(x)y_{i}}$,
$\alpha=1,\dots,m$, in presence of background $s(x)$, so ${\displaystyle f_{\alpha}(x,y_{2},\dots,x_{n})=s(x)+\tilde{f}_{\alpha}(x,y_{2},\dots,y_{n})}$
and basic statistical mapping is $F(x,\boldsymbol{y})={\displaystyle s(x)+F_{2}(x,\boldsymbol{y})}$
where 
\begin{equation}
{\displaystyle F_{2}(x,y_{2},\dots,y_{n})\doteq\ln\left(\sum_{\beta=1}^{m}e^{\tilde{f}_{\beta}(x,y_{2},\dots,y_{n})}\right)}.\label{eq: example mapping without background}
\end{equation}
One has ${\displaystyle \left(\frac{\partial F}{\partial x}\right)^{2}=(s'(x))^{2}+2\sum_{\alpha=1}^{m}s'(x)\frac{\partial\tilde{f}_{\alpha}}{\partial x}w_{\alpha}+\sum_{\alpha,\beta=1}^{m}\frac{\partial\tilde{f}_{\alpha}}{\partial x}\frac{\partial\tilde{f}_{\beta}}{\partial x}w_{\alpha}w_{\beta}}$.
Last two sums are continuous at $x=x_{0}$ since $s'(x)|_{x=x_{0}}$ is discontinuous
but finite and ${\displaystyle \left.\frac{\partial\tilde{f}_{\alpha}}{\partial x}\right|_{x=x_{0}}=0}$.
In contrast, $(s'(x))^{2}$ is discontinuous at $x=x_{0}$ since ${\displaystyle \lim_{x\rightarrow x_{0}+}(s'(x))^{2}\neq\lim_{x\rightarrow x_{0}-}(s'(x))^{2}}$.
So we have a first order phase singularity that is observed also in
$\det\boldsymbol{g}$ which is finite and discontinuous
at $x=x_{0}$. We now study the behaviour of the Gauss-Kronecker curvature near the singularity $x=x_{0}$. First, ${\displaystyle \left. \frac{\partial^{2}F_{2}}{\partial x\partial y_{i}}\right|_{x=x_{0}}=0}$ 
since ${\displaystyle \lim_{x\rightarrow x_{0}}p'_{\alpha i}(x)=0}$
for all $i=1,\dots,n$. Thus, only non-trivial term in the Hessian of $F$ near $x=x_{0}$ is ${\displaystyle \frac{\partial^{2}F}{\partial x^{2}}\mbox{Hess}[F_{2};\boldsymbol{y}]}$.
Then, $\mbox{Hess}[F;\boldsymbol{y}]=\mbox{Hess}[F_{2};\boldsymbol{y}]$ is regular and  $\mbox{Hess}[F_{2};\boldsymbol{y}]|_{x=x_{0}}=\det(A_{0})=0$
by continuity. This implies that all terms of the Hessian except ${\displaystyle s''(x)\cdot\mbox{Hess}[F_{2};\boldsymbol{y}]}$ vanish
at $x=x_{0}$. Finally,
${\displaystyle s''(x)\cdot\mbox{Hess}[F_{2};\boldsymbol{y}]}$ involves
terms ${\displaystyle \frac{x\cdot(x-x_{0})^{M}}{x_{0}\cdot\sqrt{x_{0}^{2}-x^{2}}}}$,
with $M\geq2$ and ${\displaystyle \frac{x\cdot\det(A_{0})}{x_{0}\cdot\sqrt{x_{0}^{2}-x^{2}}}\equiv 0}$. The latter can be seen in the same form with $M=+\infty$ since its singularity at $x=x_{0}$ is a removable one. All these expressions vanish at $x\rightarrow x_{0}$. So, Hessian determinant and Gauss-Kronecker curvature can be set to zero on the submanifold $\mathcal{S}:\,x=x_{0}$ where the phase singularity in ${\displaystyle \frac{\partial F}{\partial x}}$ is hidden. 

\paragraph{\label{exa: Second order phase singularity} Third example }

deals with both first and second order phase singularities. Let us take functions $\{\tilde{f}_{\alpha}(y_{2},\dots,y_{n})\}$ to define a smooth statistical
mapping $\tilde{F}$, e.g. ${\displaystyle \tilde{f}_{\alpha}(y_{2},\dots,y_{n})=\sum_{i=2}^{n}a_{\alpha i}y_{i}}$
and $\boldsymbol{a}\doteq(a_{\alpha i})_{\alpha=1,\dots,m}^{i=2,\dots,n}$.
Then consider a non-interacting background $x$ and a function $s(x)$
which expresses its energy and define $f_{\alpha}(x,\boldsymbol{y})=s(x)+\tilde{f}_{\alpha}(y_{2},\dots,y_{n})$.
Hessian determinant of $F$ is ${\displaystyle s''(x)\cdot\mbox{Hess}[\tilde{F},\boldsymbol{y}]=s''(x)\cdot(\boldsymbol{a}^{T}H\boldsymbol{a})}$.
So we can have a number of different behaviours. 
\begin{enumerate}
\item Take $s(x)=(x-x_{0})\cdot|x-x_{0}|$. So $s'(x)=2\cdot|x-x_{0}|$ and $s''(x)=2\cdot\mbox{sgn}(x-x_{0})$, which
is discontinuous and finite at $x=x_{0}$. If ${\displaystyle \mbox{Hess}[\tilde{F},\boldsymbol{y}]\equiv0}$, then  $\mbox{Hess}[F,\{x,\boldsymbol{y}\}]=0$ and the second order singularity is hidden. If instead $\mbox{Hess}(\tilde{F},\boldsymbol{y})$ does
not vanish identically, then Hessian is
discontinuous and metric determinant stays finite and continuous. Hence, the Gauss-Kronecker curvature has a jump and the second order singularity is visible.
\item Take $\tilde{F}$ such that $\mbox{Hess}[\tilde{F},\boldsymbol{y}]$ does not vanish identically. If $s(x)=\sqrt[3]{x-x_{0}}$, then at $x\sim x_{0}$ one has ${\displaystyle \mbox{Hess}(F,\boldsymbol{x})\sim(x-x_{0})^{-\frac{5}{3}}}$
and ${\displaystyle (\det\boldsymbol{g})^{\frac{n+2}{2}}\sim(x-x_{0})^{-\frac{2n+4}{3}}}$.
Thus metric determinant blows up and Gauss-Kronecker curvature
vanishes as ${\displaystyle \frac{\mbox{Hess}(F,\{x,\boldsymbol{y}\})}{(\det\boldsymbol{g})^{\frac{n+2}{2}}}\sim(x_{1}-x_{0})^{\frac{2n-1}{3}}}$. So the first order singularity is hidden on the singular locus $\mathcal{L}:\,x_{1}=x_{0}$. If instead $s(x)=\sqrt[3]{(x-x_{0})^{4}}$, then the metric determinant stays finite and continuous and the Hessian tends to $+\infty$. So Gauss-Kronecker curvature tends to $\pm\infty$ (depending on the sign of $\mbox{Hess}(\tilde{F},\boldsymbol{y}$)) as well. 
\end{enumerate}

\section{\label{sec:Tropical limit of statistical hypersurfaces} Tropical
limit }

Following the ideas of the limit-set for algebraic varieties proposed in
\cite{Bergman1971} and intensively developed later tropical geometry
(see e.g. \cite{IMS2009,Viro2011,LS2009,MS2015}),
we will study here limiting properties of statistical hypersurfaces.
Thus, we are interested in behaviour of hypersurfaces in $\mathbb{R}^{n+1}$
defined by the relation 

\begin{equation}
x_{n+1}=\ln\left(\sum_{\alpha=1}^{m}e^{f_{\alpha}(\boldsymbol{x})}\right).\label{eq: statistical mapping for tropical}
\end{equation}
at infinitely large values of the variables $x_{1},\dots,x_{n},x_{n+1}$ assuming
that they take values in unbounded intervals. Properties of statistical
hypersurfaces in such infinite blow up depend crucially on functions
$f_{\alpha}(\boldsymbol{x})$ in (\ref{eq: statistical mapping for tropical}). 

Let us begin with the simplest super-ideal case, i.e. that defined
by (\ref{eq: super ideal hypergraph}). All variables $x_{1},\dots,x_{n+1}$
enter there on equal footing. So it is natural to consider the situation when
all these variables are large uniformly, i.e. when they are given by 
\begin{equation}
x_{i}=\lambda x_{i}^{\star},\quad i=1,\dots,n,n+1\label{eq: uniform scaling}
\end{equation}
where $\lambda$ is a large parameter and all $x_{i}^{\star}$ are
finite. In terms of variables $x_{i}^{\star}$ one has a family of
super-ideal hypersurfaces $V_{n}(\lambda)$ defined by the relation
\begin{equation}
x_{n+1}^{\star}=\frac{1}{\lambda}\ln\left(\sum_{i=1}^{n}e^{\lambda x_{i}^{\star}}\right).\label{eq: super ideal case under homothety}
\end{equation}
The limiting hypersurface $V_{n}(\infty)$ ($\lambda\rightarrow\infty$)
is given by 
\begin{equation}
x_{n+1}^{\star}=\max\{x_{1}^{\star},\dots,x_{n}^{\star}\}=\sum_{i=1}^{n}\bigoplus x_{i}^{\star}\label{eq: tropical sum, super ideal case}
\end{equation}
where ${\displaystyle \sum\bigoplus}$ denotes tropical (semiring)
summation. This is the standard tropical expression (see e.g. \cite{IMS2009,Viro2011,LS2009,MS2015})
in which instead of $\lambda$ the parameter ${\displaystyle \varepsilon=\frac{1}{\lambda}}$
is usually used. 

Note that the relation (\ref{eq: uniform scaling}) can be viewed
as the homothety (uniform scaling) transformation 
\begin{equation}
x_{i}\mapsto x_{i}^{\star}=\frac{1}{\lambda}x_{i},\quad i=1,\dots,n+1. \label{eq: homothety}
\end{equation}
Hence the hypersurface given by (\ref{eq: tropical sum, super ideal case})
is the limiting ($\lambda\rightarrow\infty$) homothetic image (independent
of $\lambda$) of the hypersurface (\ref{eq: super ideal hypergraph}) for large $x_{1},\dots,x_{n+1}$. 

The tropical limit of the super-ideal statistical hypersurface
is the union of hyperplanes $P_{i}$: 
\begin{equation}
V_{n,\mbox{\ensuremath{{\scriptstyle \mathrm{trop}}}}}=\bigcup_{i=1}^{n}\overline{P}_{i}\label{eq: tropical super-ideal hypersurface}
\end{equation}
where $P_{i_{0}}=\{\vec{x}^{\star}:\ x_{n+1}^{\star}-x_{i_{0}}^{\star}=0,\,x_{i_{0}}^{\star}>x_{1}^{\star},\dots,x_{i_{0}-1}^{\star},x_{i_{0}+1}^{\star},\dots,x_{n}^{\star}\}$
and $\overline{P}_{i_{0}}$ is its closure. Note that $V_{n,\mbox{\ensuremath{{\scriptstyle \mathrm{trop}}}}}$ is the union of hyperplanes passing through the origin $(x_{i}=0,\,i=1,\dots,n+1)$. For instance, at $n=2$
$V_{2,\mbox{\ensuremath{{\scriptstyle \mathrm{trop}}}}}$ is the union
of two half-planes $\overline{P}_{1}$ and $\overline{P}_{2}$ defined
as $P_{1}=\{(x_{1}^{\star},x_{2}^{\star},x_{3}^{\star}):\ x_{3}^{\star}-x_{1}^{\star}=0,\,x_{1}^{\star}>x_{2}^{\star}\}$
and $P_{2}=\{(x_{1}^{\star},x_{2}^{\star},x_{3}^{\star}):\ x_{3}^{\star}-x_{2}^{\star}=0,\,x_{2}^{\star}>x_{1}^{\star}\}$.
Gibbs probabilities $w_{i}$ in the tropical limit takes values $0$
or $1$ on hyperplanes $P_i$, namely $w_{i_{0}}=1$ and $w_{i}=0$ if $x_{i_{0}}^{\star}>x_{i}^{\star}$
for all $i\neq i_{0}$. 

Geometric characteristics of each member of the family of hypersurfaces
(\ref{eq: super ideal case under homothety}) (at fixed $\lambda$)
is calculable directly taking into account that in terms of $x_{i}^{\star}$
the metric of the space $\mathbb{R}^{n+1}$ is ${\displaystyle \lambda^{2}((dx_{n+1}^{\star})^{2}+\sum_{i=1}^{n}(dx_{i}^{\star})^{2})}$.
So the induced metric on $V_{n}(\lambda)$ is of the form 
\begin{equation}
(ds)^{2}=\lambda^{2}\sum_{i,k=1}^{n}g_{ik}^{\star}(\lambda)dx_{i}^{\star}dx_{k}^{\star}\label{eq: uniform tropical case induced metric}
\end{equation}
where 
\begin{equation}
g_{ik}^{\star}(\lambda)=\delta_{ik}+w_{i}^{\star}(\lambda)\cdot w_{k}^{\star}(\lambda)\label{eq: uniform tropical case induced metric, coordinates}
\end{equation}
and 
\begin{equation}
w_{i}^{\star}(\lambda)=\frac{e^{\lambda x_{i}^{\star}}}{\sum_{k=1}^{n}e^{\lambda x_{k}^{\star}}}.\label{eq: uniform tropical case Gibbs weights}
\end{equation}
Similarly, for $\Omega_{ik}$, $R_{iklm}$ and Gauss-Kronecker
curvature one gets 
\begin{eqnarray}
{\displaystyle \Omega_{ik}(\lambda)=\lambda^{2}\cdot\frac{H_{ik}^{\star}(\lambda)}{\sqrt{{\displaystyle 1+\sum_{i=1}^{n}w_{i}^{\star}(\lambda)^{2}}}}},\nonumber\\
R_{iklj}(\lambda)={\displaystyle \frac{\lambda^{4}}{{\displaystyle 1+\sum_{h=1}^{n}\left(w_{h}^{\star}(\lambda)\right)^{2}}}\cdot\left[H_{kj}^{\star}(\lambda)\cdot H_{il}^{\star}(\lambda)-H_{kl}^{\star}(\lambda)\cdot H_{ij}^{\star}(\lambda)\right],}\\
{\displaystyle K(\lambda)=0}\nonumber
\label{eq: uniform tropical case, Christoffel, Riemann, Gauss-Kronecker}
\end{eqnarray}
where 
\begin{equation}
H_{ij}^{\star}(\lambda)\doteq w_{i}^{\star}(\lambda)\cdot(\delta_{ij}-w_{j}^{\star}(\lambda)).\label{eq: super ideal tropical hessian, finite part}
\end{equation}

At the limit $\lambda\rightarrow\infty$ the metric (\ref{eq: uniform tropical case induced metric, coordinates})
becomes piecewise. On each hyperplane $P_{i_{0}}$ it is a constant
diagonal one 
\begin{equation}
g_{ik,\mbox{\ensuremath{{\scriptstyle \mathrm{trop}}}}}^{(i_{0})}\equiv\lim_{\lambda\rightarrow\infty}g_{ik}^{\star}(\lambda)=\delta_{ik}(1+\delta_{i_{0}i}),\quad i,k=1,\dots n.\label{eq: tropical limit super ideal metric}
\end{equation}
On each hyperplane $P_{i_{0}}$ one also has 
\begin{eqnarray}
{\displaystyle \Omega_{ik,\mbox{\ensuremath{{\scriptstyle \mathrm{trop}}}}}\doteq\lim_{\lambda\rightarrow\infty}\frac{\Omega_{ij}(\lambda)}{\lambda^2}=0},\nonumber\\
{\displaystyle R_{iklj,\mbox{\ensuremath{{\scriptstyle \mathrm{trop}}}}}\doteq\lim_{\lambda\rightarrow\infty}\frac{R_{iklj}(\lambda)}{\lambda^{4}}=0},\\
{\displaystyle K\mbox{\ensuremath{{\scriptstyle \mathrm{trop}}}}\doteq\lim_{\lambda\rightarrow\infty}K(\lambda)=0}\nonumber
\label{eq: uniform tropical limit, Christoffel, Riemann, Gauss-Kronecker}
\end{eqnarray}
and entropy ${\displaystyle S^{(i_{0})}=\lim_{\lambda\rightarrow\infty}\lambda(x_{n+1}^{\star}-\overline{x^{\star}})=0}$.

Tropical hypersurface (\ref{eq: tropical super-ideal hypersurface})
has singularities at the points where maximum is attained on two or
more $x_{i}^{\star}$, i.e. on the hyperplanes $x_{i}^{\star}=x_{k}^{\star}$,
$x_{i}^{\star}=x_{k}^{\star}=x_{l}^{\star}$ \textit{etc}. At $n=2$
it is the line $x_{1}^{\star}=x_{2}^{\star}$. On these singularity
the derivatives ${\displaystyle {\displaystyle \frac{\partial x_{n+1}^{\star}}{\partial x_{k}^{\star}}}}$,
normal vector $\vec{N}$ and entropy $\vec{S}$ are discontinuous.
So one has first order phase singularities. On the singularities of
the type $x_{i_{0}}^{\star}=x_{k_{0}}^{\star}$ the probabilities
are ${\displaystyle w_{i_{0}}=w_{k_{0}}=\frac{1}{2}}$. On hyperplanes
$x_{i_{1}}^{\star}=x_{i_{2}}^{\star}=\dots=x_{i_{k}}^{\star}$ one
has ${\displaystyle w_{i_{1}}=w_{i_{2}}=\dots=w_{i_{k}}=\frac{1}{k}}$. 

Crossing such singular ``edges'', the metric (\ref{eq: uniform tropical case induced metric})
jumps from one diagonal to another one. On singularity edge the tropical
metric $g_{ik,{\scriptstyle \mathrm{trop}}}$ is not diagonal. For instance, on the
singularity edge $x_{i_{0}}^{\star}=x_{k_{0}}^{\star}$ one has 
\begin{equation}
g_{ij,\mbox{\ensuremath{{\scriptstyle \mathrm{trop}}}}}^{(i_{0},k_{0})}=\delta_{ij}(1+\frac{1}{4}\delta_{i_{0}i}+
\frac{1}{4}\delta_{k_{0}i})+\frac{1}{4}(\delta_{i_{0}i}\delta_{k_{0}j}+\delta_{k_{0}i}\delta_{i_{0}j})\label{eq: super ideal tropical metric edge}
\end{equation}
and 
\begin{equation}
\Omega_{ij,{\scriptstyle \mathrm{trop}}}^{(i_{0},k_{0})}=\lim_{\lambda\rightarrow\infty}\frac{\Omega_{ij}(\lambda)}{\lambda^{2}}
=\frac{{\displaystyle \delta_{ij}}\delta_{i_{0}i}+{\displaystyle \delta_{ij}}\delta_{k_{0}i}-\delta_{i_{0}j}\delta_{k_{0}i}-\delta_{k_{0}j}\delta_{i_{0}i}}{\sqrt{24}}\label{eq: super ideal tropical second fundamental form edge}
\end{equation}
for $i,j=1,\dots,n$. Christoffel symbols and curvature are also discontinuous on singularity
edges. 

The tropical limit of mean and scalar curvature is 
\begin{equation}
\Omega_{{\scriptstyle \mathrm{trop}}}\doteq\lim_{\lambda\rightarrow\infty}\Omega(\lambda){\displaystyle =\frac{1-S_{3,{\scriptstyle \mathrm{trop}}}}{\sqrt{\left(1+S_{2,{\scriptstyle \mathrm{trop}}}\right)^{3}}}}\label{eq: mean curvature tropical limit super ideal mapping}
\end{equation}
and 
\begin{equation}
{\displaystyle R_{{\scriptstyle \mathrm{trop}}}\doteq\lim_{\lambda\rightarrow\infty}R(\lambda)=\frac{2(1+S_{4,{\scriptstyle \mathrm{trop}}})}{(1+{\displaystyle S_{2,{\scriptstyle \mathrm{trop}}}})^{2}}-1}\label{eq: scalar curvature tropical limit super ideal mapping}
\end{equation}
where 
\begin{equation}
S_{p,{\scriptstyle \mathrm{trop}}}\doteq\sum_{i=1}^{n}(w_{i,{\scriptstyle \mathrm{trop}}})^{p}.\label{eq: tropical power sums}
\end{equation}
At the regular points $S_{p,{\scriptstyle \mathrm{trop}}}=1=S_{p}|_{\vec{e}_{\alpha}}$
and $\Omega_{{\scriptstyle \mathrm{trop}}}=R_{{\scriptstyle \mathrm{trop}}}=0$.
At the singularity edge with $x_{i_{1}}^{\star}=x_{i_{2}}^{\star}=\dots=x_{i_{r}}^{\star}$
one has ${\displaystyle S_{p,{\scriptstyle \mathrm{trop}}}=\frac{1}{r^{p-1}}}$
and, hence 
\begin{equation}
\Omega_{{\scriptstyle \mathrm{trop}}}^{(r)}=\frac{r-1}{\sqrt{r(r+1)}},\quad R_{{\scriptstyle \mathrm{trop}}}^{(r)}=\frac{(r-1)(r-2)}{r(r+1)}.\label{eq: tropical mean scalar curvatures r-edge}
\end{equation}
At the most singular edge with $r=n$ 
\begin{equation}
\Omega_{{\scriptstyle \mathrm{trop}}}^{(n)}=\frac{n-1}{\sqrt{n(n+1)}},\quad R_{{\scriptstyle \mathrm{trop}}}^{(n)}=\frac{(n-1)(n-2)}{n(n+1)}\label{eq: tropical mean scalar curvatures n-edge}
\end{equation}
that coincide with maximum values of mean and scalar curvatures of
super-ideal statistical hypersurface.

Note that the tropical limit in statistical physics of macroscopic
systems with highly degenerate energy levels studied in \cite{AK2015}
corresponds to a very special, essentially one-dimensional case of
above consideration when all ${\displaystyle x_{i}=S_{i}-\frac{\varepsilon_{i}}{T}}$, $S_{i}$, $\varepsilon_{i}$ are constants and $T$ is a variable
(temperature). Scaling parameter used in \cite{AK2015}
is ${\displaystyle \lambda=\frac{1}{k}}$ where $k$ is the Boltzmann
constant. So, the results obtained in \cite{AK2015}
describe some properties of line sections of the tropical limit of
super-ideal statistical hypersurfaces. 

Tropical limit of ideal hypersurfaces given by (\ref{eq: linear model})
is formally quite similar to the super-ideal case. It is given by
\begin{equation}
x_{n+1}^{\star}=\max\left\{ \sum_{i=1}^{n}a_{\alpha i}x_{i}^{\star},\ \alpha=1,\dots,m\right\} =\sum_{\alpha=1}^{m}\bigoplus(\sum_{i=1}^{n}a_{\alpha i}x_{i}^{\star}).\label{eq: tropical sum, linear case}
\end{equation}
However, the presence of parameters $a_{\alpha i}$ and the fact that
$n\neq m$ make the situation richer. Tropical ideal hypersurface
is the union of $m$ hyperplanes $P_{\alpha}$, namely 
\begin{equation}
V_{n,\mbox{\ensuremath{{\scriptstyle \mathrm{trop}}}}}^{\mbox{\ensuremath{{\scriptstyle \mathrm{linear}}}}}=\bigcup_{\alpha=1}^{m}\overline{P}_{\alpha}\label{eq: tropical linear hypersurface}
\end{equation}
where we have defined ${\displaystyle P_{\alpha}=\{\overrightarrow{x^{\star}}:\ x_{n+1}^{\star}=\sum_{i=1}^{n}a_{\alpha i}x_{i}^{\star},\,\sum_{i=1}^{n}a_{\alpha i}x_{i}^{\star}>\sum_{i=1}^{n}a_{\beta i}x_{i}^{\star},\,\beta\neq\alpha\}}$.
Outside the singular sector, metric (\ref{eq: metric in linear case})
on $P_{\alpha}$ is again equal to a constant metric depending on
$a_{\alpha i}$, i.e. 
\begin{equation}
g_{ij,\mbox{\ensuremath{{\scriptstyle \mathrm{trop}}}}}^{(\alpha)}\doteq\lim_{\lambda\rightarrow\infty}g_{ij}^{\star}=\delta_{ij}+a_{\alpha i}\cdot a_{\alpha j},\quad i,j=1,\dots n.\label{eq: linear tropical metric generic}
\end{equation}
Depending on $a_{\alpha i}$ there are variety of singularity hyperplanes
on which metric, normal vector and entropy are discontinuous having
specific values on singularity edges. 

At the special case discussed at the end of the section \ref{sec: Ideal statistical hypersurfaces } tropical limit
considered above is closely connected with the tropical limit of $\log\mbox{\ensuremath{\tau}}$ for $m$-soliton solutions for Korteweg\textendash de Vries and Kadomtsev\textendash Petviashvili equations studied in \cite{BC2006,CK2008,DH-H2011,DH-H2014}.

\section{\label{sec: Double scaling tropical limit in non-ideal case} Double
scaling tropical limit in non-ideal case }

 Tropical limit of non-ideal statistical hypersurfaces is more complicated
due to the variety of possible behaviour of functions $f_{\alpha}(\boldsymbol{x})$
under dilatation. In a simple case, when all functions $f_{\alpha}(\boldsymbol{x})$
are homogeneous functions of degree one ($f_{\alpha}(\lambda\boldsymbol{x^{\star}})=\lambda f_{\alpha}(\boldsymbol{x^{\star}})$), the corresponding statistical
hypersurface in the standard tropical limit is given by 
\begin{equation}
x_{n+1}^{\star}=\max\left\{ f_{1}(\boldsymbol{x^{\star}}),f_{2}(\boldsymbol{x^{\star}}),\dots,f_{m}(\boldsymbol{x^{\star}})\right\} =\sum_{\alpha=1}^{m}\bigoplus f_{\alpha}(\boldsymbol{x^{\star}}).\label{eq: tropical sum, homogeneous d=1 case}
\end{equation}

So 
\begin{equation}
V_{n,\mbox{\ensuremath{{\scriptstyle \mathrm{trop}}}}}^{\mbox{\mbox{\ensuremath{{\scriptstyle \mathrm{non-ideal}}}}}}=\bigcup_{\alpha=1}^{m}\overline{V}_{\alpha}\label{eq: tropical non-ideal hypersurface}
\end{equation}
where the hypersurface $V_{\alpha_{0}}$ is defined by 
\begin{equation}
{\displaystyle V_{\alpha_{0}}=\{\overrightarrow{x^{\star}}:\ x_{n+1}^{\star}-f_{\alpha_{0}}(\boldsymbol{x^{\star}})=0,\,f_{\alpha_{0}}(\boldsymbol{x^{\star}})>f_{\alpha}(\boldsymbol{x^{\star}}),\,\alpha\neq\alpha_{0}\}}\label{eq: piece tropical non-ideal hypersurface}
\end{equation}
and $\overline{V}_{\alpha_{0}}$ is its closure. On the hypersurface
$V_{\alpha_{0}}$ the probability $w_{\alpha_{0}}=1$ while $w_{\beta}=0$,
$\beta\neq\alpha_{0}$. 

Hence, the tropical limit of the metric (\ref{eq: basic metric 1})
on $V_{\alpha_{0}}$ is 
\begin{equation}
g_{ik,\mbox{\ensuremath{{\scriptstyle \mathrm{trop}}}}}^{(\alpha_{0})}=\delta_{ik}+\frac{\partial f_{\alpha_{0}}}{\partial x_{i}^{\star}}\frac{\partial f_{\alpha_{0}}}{\partial x_{k}^{\star}},\quad i,k=1,\dots n;\label{eq: tropical non-ideal metric generic}
\end{equation}
there is no summation on $\alpha_{0}$ here. Then, on each $V_{\alpha_{0}}$
one has tropical limits of $\Gamma_{ik}^{l}$, $\Omega_{ik}$, $R_{iklj}$ and $K$
given by formulae (\ref{eq: Christoffel symbols})-(\ref{eq: Gauss-Kronecker curvature})
in which instead of summation over $\alpha$ there is tropical summation
over $\alpha$, i.e. there is only the term with $\alpha=\alpha_{0}$
since ${\displaystyle \bar{f}_{i}=\frac{\partial f_{\alpha_{0}}}{\partial x_{i}^{\star}}}$.
The tropical limit of the entropy on $V_{\alpha_{0}}$ is ${\displaystyle S_{\alpha_{0}}=x_{n+1}^{\star}-f_{\alpha_{0}}(\boldsymbol{x^{\star}})=0}$.
Singularity ``edges'' now are hypersurfaces of the type $f_{\alpha_{0}}(\boldsymbol{x^{\star}})=f_{\alpha_{1}}(\boldsymbol{x^{\star}})$
on which all characteristics are discontinuous. 

Situation is quite different in the case when functions $f_{\alpha}(\boldsymbol{x})$
are all homogeneous of degree $d>1$. In such a case the definition
(\ref{eq: basic statistical mapping}) implies that in the tropical
regime the variables $x_{1},\dots,x_{n},x_{n+1}$ are large, but not
uniformly. The natural parametrization of large variables, instead
of (\ref{eq: uniform scaling}), now is 
\begin{equation}
x_{i}=\lambda x_{i}^{\star},\,i=1,\dots,n;\quad x_{n+1}=\lambda^{d}x_{n+1}^{\star}.\label{eq: double scaling}
\end{equation}
It is easy to see that only with such a rescaling the hypersurface
$V_{n}(\lambda)$ defined by 
\begin{equation}
x_{n+1}^{\star}=\frac{1}{\lambda^{d}}\ln\left(\sum_{\alpha=1}^{m}e^{\lambda^{d}f_{\alpha}(\boldsymbol{x^{\star})}}\right)\label{eq: non-ideal tropical mapping homogeneous degree d double scaling}
\end{equation}
has finite, independent on $\lambda$, tropical limit at $\lambda\rightarrow\infty$.
It is given by the formula 
\begin{equation}
x_{n+1}^{\star}=\max\{f_{1}(\boldsymbol{x^{\star}}),f_{2}(\boldsymbol{x^{\star}}),\dots,f_{m}(\boldsymbol{x^{\star}})\}=\sum_{\alpha=1}^{m}\bigoplus f_{\alpha}(\boldsymbol{x^{\star}}).\label{eq: tropical limit homogeneous degree d}
\end{equation}

The squared line element of the space $\mathbb{R}^{n+1}$ under this
rescaling becomes 
\begin{equation}
(ds)^{2}=\lambda^{2d}(dx_{n+1}^{\star})^{2}+\lambda^{2}\cdot\sum_{i=1}^{n}(dx_{i}^{\star})^{2}.\label{eq: tropical limit metric ambient space degree d}
\end{equation}
So the induced metric of the hypersurface $V_{n}$ (\ref{eq: non-ideal tropical mapping homogeneous degree d double scaling}) is of the form 
\begin{equation}
g_{ik}(\lambda)=\lambda^{2}\delta_{ik}+\lambda^{2d}\cdot\overline{f_{i}(\lambda)}\cdot\overline{f_{k}(\lambda)},\quad i,k=1,\dots,n\label{eq: tropical limit induced metric degree d}
\end{equation}
where 
\begin{equation}
\bar{f}_{i}(\lambda)\doteq\sum_{\alpha=1}^{m}w_{\alpha}(\lambda)\cdot\frac{\partial f_{\alpha}}{\partial x_{i}^{\star}}\label{eq: tropical limit first derivatives degree d}
\end{equation}
and unit normal vector (with respect to the metric (\ref{eq: tropical limit metric ambient space degree d})) is 
\begin{equation}
\vec{N}(\lambda)=\frac{\lambda^{n+d-2}}{\sqrt{\det\boldsymbol{g}(\lambda)}}
\left(-\bar{f}_{1}(\lambda),-\bar{f}_{2}(\lambda),\dots,-\bar{f}_{n}(\lambda),
\lambda^{2-2d}\right)
\label{eq: tropical limit unit normal vector degree d}
\end{equation}
Hence, in the limit $\lambda\rightarrow\infty$, on each hypersurface
$V_{\alpha_{0}}$ one has 
\begin{equation}
g_{ik,\mbox{\ensuremath{{\scriptstyle \mathrm{trop}}}}}^{(\alpha_{0})}\doteq\lim_{\lambda\rightarrow\infty}\frac{g_{ij}(\lambda)}{\lambda^{2d}}=\frac{\partial f_{\alpha_{0}}}{\partial x_{i}^{\star}}\cdot\frac{\partial f_{\alpha_{0}}}{\partial x_{k}^{\star}},\quad i,k=1,\dots,n.\label{eq: tropical limit induced metric degree d, regular sectors}
\end{equation}
At large $\lambda$ the dominant terms in $\Gamma_{ik}^{l}(\lambda)$, $\Omega_{ik}(\lambda)$, $R_{iklm}(\lambda)$ and $K(\lambda)$ are of the orders $0$, $1$, $2$ and $2-n-2d$ in $\lambda$, respectively. Hence, on $V_{\alpha_{0}}$ 
\begin{equation}
\Gamma_{ik,{\scriptstyle \mathrm{trop}}}^{l\,(\alpha_{0})}\doteq\lim_{\lambda\rightarrow\infty}\Gamma_{ik}^{l}(\lambda)=\frac{{\displaystyle \frac{\partial^{2}f_{\alpha_{0}}}{\partial x_{i}^{\star}\partial x_{j}^{\star}}\cdot\frac{\partial f_{\alpha_{0}}}{\partial x_{l}^{\star}}}}{{\displaystyle \sum_{h=1}^{n}\left(\frac{\partial f_{\alpha_{0}}}{\partial x_{h}^{\star}}\right)^{2}}},\label{eq: non-uniform, degree d tropical case, Christoffel}
\end{equation}
\begin{equation}
\Omega_{ij,{\scriptstyle \mathrm{trop}}}^{(\alpha_{0})}\doteq\lim_{\lambda\rightarrow\infty}\frac{\Omega_{ij}(\lambda)}{\lambda}=\frac{{\displaystyle \frac{\partial f_{\alpha_{0}}}{\partial x_{i}^{\star}\partial x_{j}^{\star}}}}{{\displaystyle \sqrt{{\displaystyle \sum_{h=1}^{n}\left(\frac{\partial f_{\alpha_{0}}}{\partial x_{h}^{\star}}\right)^{2}}}}},
\label{eq: non-uniform, degree d tropical case, second fundamental form}
\end{equation}
\begin{eqnarray}
R_{iklj,{\scriptstyle \mathrm{trop}}}^{(\alpha_{0})}&\doteq\lim_{\lambda\rightarrow\infty}\frac{R_{iklj}(\lambda)}{\lambda^{2}}\nonumber \\&=\frac{1}{{\displaystyle \sum_{h=1}^{n}\left(\frac{\partial f_{\alpha_{0}}}{\partial x_{h}^{\star}}\right)^{2}}}\cdot\left[\frac{\partial^{2}f_{\alpha_{0}}}{\partial x_{k}^{\star}\partial x_{j}^{\star}}\cdot\frac{\partial^{2}f_{\alpha_{0}}}{\partial x_{i}^{\star}\partial x_{l}^{\star}}-\frac{\partial^{2}f_{\alpha_{0}}}{\partial x_{k}^{\star}\partial x_{l}^{\star}}\cdot\frac{\partial^{2}f_{\alpha_{0}}}{\partial x_{i}^{\star}\partial x_{j}^{\star}}\right]\label{eq: non-uniform, degree d tropical case, Riemann}
\end{eqnarray}
and 
\begin{equation}
K_{{\scriptstyle \mathrm{trop}}}^{(\alpha_{0})}\doteq\lim_{\lambda\rightarrow\infty}\lambda^{n+2d-2}\cdot K(\lambda)=\frac{\det\left|{\displaystyle \frac{\partial^{2}f_{\alpha_{0}}}{\partial x_{i}^{\star}\partial x_{j}^{\star}}}\right|}{\left[{\displaystyle \sum_{h=1}^{n}\left(\frac{\partial f_{\alpha_{0}}}{\partial x_{h}^{\star}}\right)^{2}}\right]^{\frac{n}{2}+1}}.\label{eq: non-uniform, degree d tropical case, Gauss-Kronecker}
\end{equation}

Dominant behaviours at the limit $\lambda\rightarrow\infty$ changes drastically
on singular locus where two or more $f_{\alpha_{1}}=f_{\alpha_{2}}=\dots=f_{\alpha_{r}}$
attain the maximum. Some results are discussed in the Appendix \ref{sec: Appendix B }. 

Tropical metric (\ref{eq: tropical limit induced metric degree d, regular sectors})
is degenerate. It is a consequence of the degeneration of the metric
(\ref{eq: tropical limit metric ambient space degree d}) in $\mathbb{R}^{n+1}$.
Consequently, the tropical limit of other geometric characteristics
has a rather special structure too. 

We see that in this case the double scaling limit defined via (\ref{eq: double scaling}) provides us with the effective tropical limit, in constrast to
the usual scaling limit. Note that the double scaling limit technique
is a widely used tool in statistical physics and quantum field theory
(see e.g. \cite{DGZ-J1995}). 

The double scaling tropical limit is useful also in cases of more
general functions $f_{\alpha}(\boldsymbol{x})$. For instance, if
\begin{equation}
f_{\alpha}(\boldsymbol{x})=\sum_{i=1}^{n}a_{\alpha i}x_{i}+\varphi_{\alpha}(\boldsymbol{x})\label{eq: non homogeneous function}
\end{equation}
where all $\varphi_{\alpha}(\boldsymbol{x})$ are homogeneous functions
of degree $d>1$, then the limit (\ref{eq: double scaling}) produces
the tropical hypersurface given by 

\begin{equation}
x_{n+1}^{\star}=\max\{\varphi_{1}(\boldsymbol{x^{\star}}),\dots,\varphi_{m}(\boldsymbol{x^{\star}})\}=\sum_{\alpha=1}^{m}\bigoplus\varphi_{\alpha}(\boldsymbol{x^{\star}}).\label{eq: linear plus interactions tropical mapping}
\end{equation}
In this case, the tropical limit is defined by the nonlinear (interaction)
terms. 

A simple example is provided by the hypersurface in (\ref{eq: interacting, example 2, mapping})
with $d=2$, $m=2$, $n=3$ and $\varepsilon >0$. The double rescaling now is 
\begin{equation}
x_{i}=\lambda x_{i}^{\star},\,i=1,2,3;\quad x_{4}=\lambda^{2}x_{4}^{\star}.\label{eq: double scaling, example}
\end{equation}
and the double scaling tropical limit of the hypersurface (\ref{eq: interacting, example 2, mapping})
is given by 
\begin{equation}
x_{4}^{\star}=\varepsilon\cdot\max\{x_{1}^{\star}x_{2}^{\star},x_{1}^{\star}x_{3}^{\star}\}=\varepsilon(x_{1}^{\star}x_{2}^{\star})\oplus\varepsilon(x_{1}^{\star}x_{3}^{\star}).\label{eq: tropical limit degree 2 example}
\end{equation}
It is the union of hypersurfaces 
\begin{equation}
V_{3,\mbox{\ensuremath{{\scriptstyle \mathrm{trop}}}}}=\overline{V}_{1}\cup\overline{V}_{2}\label{eq: tropical non-ideal hypersurface, example}
\end{equation}
where 
\begin{eqnarray}
{\displaystyle V_{1}=\{(x_{1}^{\star},x_{2}^{\star},x_{3}^{\star},x_{4}^{\star}):\ x_{4}^{\star}-\varepsilon x_{1}^{\star}x_{2}^{\star}=0,\,\varepsilon x_{1}^{\star}x_{2}^{\star}>\varepsilon x_{1}^{\star}x_{3}^{\star}\}}\nonumber\\
{\displaystyle V_{2}=\{(x_{1}^{\star},x_{2}^{\star},x_{3}^{\star},x_{4}^{\star}):\ x_{4}^{\star}-\varepsilon x_{1}^{\star}x_{3}^{\star}=0,\,\varepsilon x_{1}^{\star}x_{3}^{\star}>\varepsilon x_{1}^{\star}x_{2}^{\star}\}}.
\label{eq: piece tropical non-ideal hypersurface, example}
\end{eqnarray}
On $V_{1}$ the tropical metric and second fundamental form are  
\begin{equation}
g_{ik,\mbox{\ensuremath{{\scriptstyle \mathrm{trop}}}}}^{(1)}=\varepsilon^{2}
\left(\begin{array}{*{3}{c}}x_{2}^{\star2} & x_{1}^{\star}x_{2}^{\star} & 0\\
x_{1}^{\star}x_{2}^{\star} & x_{1}^{\star2} & 0\\
0 & 0 & 0
\end{array}\right),\quad{\displaystyle \Omega_{ij,{\scriptstyle \mathrm{trop}}}^{(1)}=\frac{\delta_{i1}\cdot\delta_{j2}+\delta_{i2}\cdot\delta_{j1}}{{\displaystyle \sqrt{x_{1}^{2}+x_{2}^{2}}}}}\label{eq: tropical non-ideal hypersurface, example, metric on piece 1}
\end{equation}
and on $V_{2}$ 
\begin{equation}
g_{ik,\mbox{\ensuremath{{\scriptstyle \mathrm{trop}}}}}^{(2)}=\varepsilon^{2}\left(\begin{array}{*{3}{c}}x_{3}^{\star2} & 0 & x_{1}^{\star}x_{3}^{\star}\\
0 & 0 & 0\\
x_{1}^{\star}x_{3}^{\star} & 0 & x_{1}^{\star2}
\end{array}\right),\quad{\displaystyle \Omega_{ij,{\scriptstyle \mathrm{trop}}}^{(2)}=\frac{\delta_{i1}\cdot\delta_{j3}+\delta_{i3}\cdot\delta_{j1}}{{\displaystyle \sqrt{x_{1}^{2}+x_{3}^{2}}}}}.\label{eq: tropical non-ideal hypersurface, example, metric on piece 2}
\end{equation}
One also has  
\begin{equation}
{\displaystyle R_{iklj,{\scriptstyle \mathrm{trop}}}^{(1)}=\frac{1}{{\displaystyle x_{1}^{\star2}+x_{2}^{\star2}}}\cdot\left\{ \begin{array}{l}
1,\quad\mbox{if }k=l=1,i=j=2\\
1,\quad\mbox{if }k=l=2,i=j=1\\
-1,\quad\mbox{if }k=j=1,i=l=2\\
-1,\quad\mbox{if }k=j=2,i=l=1\\
0,\quad\mbox{otherwise}
\end{array}\right.},\quad K_{\mbox{\ensuremath{{\scriptstyle \mathrm{trop}}}}}^{(1)}=0
\label{eq: non homogeneous tropical limit, Riemann, Gauss-Kronecker, example, piece 1}
\end{equation}
and 
\begin{equation}
{\displaystyle R_{iklj,{\scriptstyle \mathrm{trop}}}^{(2)}=\frac{1}{{\displaystyle x_{1}^{\star2}+x_{3}^{\star2}}}\cdot\left\{ \begin{array}{l}
1,\quad\mbox{if }k=l=1,i=j=3\\
1,\quad\mbox{if }k=l=3,i=j=1\\
-1,\quad\mbox{if }k=j=1,i=l=3\\
-1,\quad\mbox{if }k=j=3,i=l=1\\
0,\quad\mbox{otherwise}
\end{array}\right.},\quad K_{\mbox{\ensuremath{{\scriptstyle \mathrm{trop}}}}}^{(2)}=0.\label{eq: non homogeneous tropical limit, Riemann, Gauss-Kronecker, example, piece 2}
\end{equation}
Comparing the ideal and non-ideal cases we see that in the tropical
limit difference between them becomes easily visible geometrically.
Indeed, the tropical limit of ideal statistical hypersurface is a
piecewise hyperplane while in the non-ideal case it is piecewise
curved hypersurface. Moreover, the double scaling tropical limit reveals
the dominant role of interactions (nonlinear terms). 

In more details the double and multi-scaling versions of the tropical
limit and their applications to statistical physics, study of statistical
hypersurfaces and other geometric objects will be considered in a
separate publication. 

\appendix

\section{\label{sec: Appendix A }}

Here we will use, for notational simplicity, both double index notation
$x_{i}^{p}$ and single index notation $x_{i}$, with $i=1,2,\dots,n$,
corresponding to the ordering of coordinates first by index $p$ and then
by index $i$: $(x_{1},x_{2},\dots,x_{n})=(x_{1}^{1},\dots,x_{q_{1}}^{1},x_{1}^{2},\dots,x_{q_{2}}^{2},x_{1}^{3},\dots,x_{q_{P}-1}^{P},x_{q_{P}}^{P})$. 
\\
\textit{Proof: }
The first order correction for the determinant ${\displaystyle \det C={\displaystyle \sum_{\sigma\in\mathcal{S}(n)}\mbox{sgn}(\sigma)\cdot\prod_{i=1}^{n}c_{i\sigma(i)}}}$
of any matrix 
${\displaystyle C=\left(c_{ij}\right)_{1\leq i,j\leq n}}$ whose entries
depends on a parameter $\varepsilon$ is given by sum of first order
corrections for each term: 
\begin{equation}
I_{\varepsilon}[\det A]=\sum_{\sigma\in\mathcal{S}(n)}\mbox{sgn}(\sigma)\cdot\sum_{j=1}^{n}I_{\varepsilon}[c_{j\sigma(j)}]\prod_{j\neq i=1}^{n}c_{i\sigma(i)}|_{\varepsilon=0}.
\label{eq: first order correction determinant of a matrix}
\end{equation}
Applying this to the Hessian determinant of $F$, we get 
\begin{equation}
\sum_{\sigma\in\mathcal{S}(n)}\mbox{sgn}(\sigma)\cdot\sum_{j=1}^{n}I_{\varepsilon}[\frac{\partial^{2}F}{\partial x_{j}\partial x_{\sigma(j)}}]\cdot\prod_{j\neq i=1}^{n}\frac{\partial^{2}F|_{\varepsilon=0}}{\partial x_{i}\partial x_{\sigma(i)}}.
\label{eq: first order correction determinant Hessian statistical mapping }
\end{equation}
It follows from (\ref{eq: ideal p-subsystems mapping}) that each product ${\displaystyle \prod_{j\neq i=1}^{n}\frac{\partial^{2}F|_{\varepsilon=0}}{\partial x_{i}\partial x_{\sigma(i)}}}$
is non-vanishing only when both $i$ and $\sigma(i)$ belong to the
same subsystem $p$, for all $i\neq j$. In such a case also $j$
and $\sigma(j)$ must belong to the same subsystem, say $p_{j}$, as
follows from injectivity of $\sigma$. So, non-vanishing terms correspond
to ${\displaystyle \sigma\in\prod_{p=1}^{P}\mathcal{S}(q_{p})}$.
Now we can rewrite the first order correction of Hessian determinant
as 
\begin{eqnarray}
&\sum_{\sigma_{1}\in\mathcal{S}(q_{1})}\cdots\sum_{\sigma_{P}\in\mathcal{S}(q_{P})}
\mbox{sgn}(\sigma_{1})\cdot\dots\mbox{sgn}(\sigma_{P})\cdot\sum_{j=1}^{n}I_{\varepsilon}[\frac{\partial^{2}F}{\partial x_{j}\partial x_{\sigma(j)}}]\cdot\prod_{j\neq i=1}^{n}\frac{\partial^{2}F|_{\varepsilon=0}}{\partial x_{i}\partial x_{\sigma(i)}}\nonumber \\&=\sum_{p=1}^{P}\det(1,p,F)\cdot\prod_{p\neq r=1}^{P}\det(0,r,F)\label{eq: formula first order correction determinant}
\end{eqnarray}
where we have defined 
\begin{equation}
\det(0,p,F)\doteq{\displaystyle \sum_{\sigma_{p}\in\mathcal{S}(q_{p})}\mbox{sgn}(\sigma_{p})\cdot\prod_{i=1}^{q_{p}}\frac{\partial^{2}F|_{\varepsilon=0}}{\partial x_{i}^{p}\partial x_{\sigma_{p}(i)}^{p}}}\label{eq: partial determinants zeroth order}
\end{equation}
 and 
\begin{equation}
{\displaystyle \det(1,p,F)\doteq \sum_{\sigma_{p}\in\mathcal{S}(q_{p})}\sum_{j=1}^{q_{p}}\mbox{sgn}(\sigma_{p})\cdot I_{\varepsilon}[\frac{\partial^{2}F}{\partial x_{j}^{p}\partial x_{\sigma_{p}(j)}^{p}}]\cdot\prod_{j\neq i=1}^{q_{p}}\frac{\partial^{2}F|_{\varepsilon=0}}{\partial x_{i}^{p}\partial x_{\sigma_{p}(i)}^{p}}.}\label{eq: partial determinants first order}
\end{equation}
Each term in ${\displaystyle \sum_{p=1}^{P}\det(1,p,F)\cdot\prod_{p\neq r=1}^{P}\det(0,r,F)}$
contains a factor $\det(0,r,F)$, which is equal to zero since it is the Hessian
determinant of a system of the form (\ref{eq: partial statistical functions}). So the whole sum vanishes. Hence, the Hessian of $F$ is $\mathcal{O}(\varepsilon^{2})$,
i.e. its first order correction is equal to zero. The series expansion of ${\displaystyle (\det\boldsymbol{g})^{-\frac{n+2}{2}}}$
is regular at $\varepsilon=0$. Hence first non vanishing term in
the expansion of Gauss-Kronecker curvature is at least of second order
in $\varepsilon$.
\hfill\ensuremath{\square}

\section{\label{sec: Appendix B }}

The study of statistical hypersurface (\ref{eq: non-ideal tropical mapping homogeneous degree d double scaling})
for large $\lambda$ shows major differences between regular sector
$V_{\alpha}$ and singular sector $\overline{V}_{\alpha}\backslash V_{\alpha}$,
$\alpha=1,\dots,m$. We recall that the former is the set where ${\displaystyle \max_{\alpha}\{f_{\alpha}(\boldsymbol{x^{\star}})\}}$
is attained only once, the latter is the set where the maximum is
attained at least twice. In the following we suppose that $\{\bar{\alpha}_{1},\dots,\bar{\alpha}_{r}\}$
is the subset of indices $\{1,2,\dots,m\}$ where ${\displaystyle \max_{\alpha}\{f_{\alpha}(\boldsymbol{x})\}}$
is attained. On the singular sector $r>1$ and one has 
\begin{equation}
{\displaystyle {\displaystyle w_{\alpha,\mbox{\ensuremath{{\scriptstyle \mathrm{trop}}}}}}=\lim_{\lambda\rightarrow\infty}w_{\alpha}(\lambda)=\frac{1}{r}\cdot\sum_{p=1}^{r}\delta_{\bar{\alpha}_{p}\alpha}},\label{eq: strong tropical Gibbs weights}
\end{equation}
\begin{equation}
\bar{f}_{i,\mbox{\ensuremath{{\scriptstyle \mathrm{trop}}}}}={\displaystyle \lim_{\lambda\rightarrow\infty}\frac{\partial F(\lambda)}{\partial x_{i}^{\star}}=\frac{1}{r}\cdot\sum_{p=1}^{r}\frac{\partial f_{\bar{\alpha}_{p}}}{\partial x_{i}^{\star}}},\label{eq: strong tropical gradient mapping}
\end{equation}
and 
\begin{equation}
{\displaystyle \Phi_{ij,\mbox{\ensuremath{{\scriptstyle \mathrm{trop}}}}}\doteq\lim_{\lambda\rightarrow\infty}\frac{1}{\lambda^{d}}\frac{\partial F(\lambda)}{\partial x_{i}^{\star}\partial x_{j}^{\star}}=\frac{1}{r}\cdot\sum_{p=1}^{r}\frac{\partial f_{\bar{\alpha}_{p}}}{\partial x_{i}^{\star}}\frac{\partial f_{\bar{\alpha}_{p}}}{\partial x_{j}^{\star}}-\bar{f}_{i,\mbox{\ensuremath{{\scriptstyle \mathrm{trop}}}}}\cdot\bar{f}_{j,\mbox{\ensuremath{{\scriptstyle \mathrm{trop}}}}}}.\label{eq: strong tropical hessian}
\end{equation}
After simple computations, one finds 
\begin{equation}
{\displaystyle g_{ij,\mbox{\ensuremath{{\scriptstyle \mathrm{trop}}}}}=\lim_{\lambda\rightarrow\infty}\frac{g_{ij}(\lambda)}{\lambda^{2d}}=\bar{f}_{i,\mbox{\ensuremath{{\scriptstyle \mathrm{trop}}}}}\cdot\bar{f}_{j,\mbox{\ensuremath{{\scriptstyle \mathrm{trop}}}}},}\label{eq: strong tropical metric}
\end{equation}
\begin{equation}
{\displaystyle \det\boldsymbol{g}_{\mbox{\ensuremath{{\scriptstyle \mathrm{trop}}}}}=\lim_{\lambda\rightarrow\infty}\frac{\det\boldsymbol{g}(\lambda)}{\lambda^{2n+2d-2}}=\delta_{1,d}+\sum_{i=1}^{n}\left(\bar{f}_{i,\mbox{\ensuremath{{\scriptstyle \mathrm{trop}}}}}\right)^{2}},\label{eq: strong tropical determinant metric}
\end{equation}
\begin{eqnarray}
\Gamma_{ij,{\scriptstyle \mathrm{trop}}}^{l}&{\displaystyle \doteq\lim_{\lambda\rightarrow\infty}\frac{\Gamma_{ij}^{l}(\lambda)}{\lambda^{d}}=\frac{\Phi_{ij,\mbox{\ensuremath{{\scriptstyle \mathrm{trop}}}}}\cdot\bar{f}_{l,\mbox{\ensuremath{{\scriptstyle \mathrm{trop}}}}}}{\det\boldsymbol{g}_{\mbox{\ensuremath{{\scriptstyle \mathrm{trop}}}}}}}\nonumber\\&{\displaystyle=\frac{{\displaystyle \frac{\bar{f}_{l,\mbox{\mbox{\ensuremath{{\scriptstyle \mathrm{trop}}}}}}}{r}\cdot\sum_{p=1}^{r}\frac{\partial f_{\bar{\alpha}_{p}}}{\partial x_{i}^{\star}}\frac{\partial f_{\bar{\alpha}_{p}}}{\partial x_{j}^{\star}}-\bar{f}_{i,\mbox{\ensuremath{{\scriptstyle \mathrm{trop}}}}}\cdot\bar{f}_{j,\mbox{\ensuremath{{\scriptstyle \mathrm{trop}}}}}\cdot\bar{f}_{l,\mbox{\ensuremath{{\scriptstyle \mathrm{trop}}}}}}}{\displaystyle{\delta_{1,d}+\sum_{k=1}^{n}\left(\bar{f}_{k,\mbox{\ensuremath{{\scriptstyle \mathrm{trop}}}}}\right)^{2}}}},\label{eq: strong tropical Christoffel}
\end{eqnarray}
\begin{equation}
\Omega_{ij,{\scriptstyle \mathrm{trop}}}\doteq\lim_{\lambda\rightarrow\infty}\frac{\Omega_{ij}(\lambda)}{\lambda^{d+1}}=\frac{{\displaystyle \frac{1}{r}\sum_{p=1}^{r}\frac{\partial f_{\bar{\alpha}_{p}}}{\partial x_{i}^{\star}}\frac{\partial f_{\bar{\alpha}_{p}}}{\partial x_{j}^{\star}}-\bar{f}_{i,{\scriptstyle \mathrm{trop}}}\cdot\bar{f}_{j,{\scriptstyle \mathrm{trop}}}}}{{\displaystyle \sqrt{{\displaystyle \delta_{1,d}+\sum_{h=1}^{n}\left(\bar{f}_{h,{\scriptstyle \mathrm{trop}}}\right)^{2}}}}},\label{eq: strong tropical second fundamental form}
\end{equation}
\begin{equation}
R_{iklj,{\scriptstyle \mathrm{trop}}}\doteq\lim_{\lambda\rightarrow\infty}\frac{R_{iklj}(\lambda)}{\lambda^{2d+2}}=\frac{\Phi_{kj,\mbox{\ensuremath{{\scriptstyle \mathrm{trop}}}}}\cdot\Phi_{il,\mbox{\ensuremath{{\scriptstyle \mathrm{trop}}}}}-\Phi_{kl,\mbox{\ensuremath{{\scriptstyle \mathrm{trop}}}}}\cdot\Phi_{ij,\mbox{\ensuremath{{\scriptstyle \mathrm{trop}}}}}}{\displaystyle{\delta_{1,d}+\sum_{h=1}^{n}\left(\bar{f}_{h,\mbox{\ensuremath{{\scriptstyle \mathrm{trop}}}}}\right)^{2}}}\label{eq: strong tropical Riemann curvature tensor}
\end{equation}
and 
\begin{eqnarray}
K_{{\scriptstyle \mathrm{trop}}}&\doteq\lim_{\lambda\rightarrow\infty}\lambda^{n+2d-2-dn}\cdot K(\lambda)\nonumber\\&=\frac{\det\left|{\displaystyle \frac{1}{r}\sum_{p=1}^{r}\frac{\partial f_{\bar{\alpha}_{p}}}{\partial x_{i}^{\star}}\frac{\partial f_{\bar{\alpha}_{p}}}{\partial x_{j}^{\star}}-\bar{f}_{i,{\scriptstyle \mathrm{trop}}}\cdot\bar{f}_{j,{\scriptstyle \mathrm{trop}}}}\right|}{\left[{\displaystyle \delta_{1,d}+\sum_{i=1}^{n}\left(\bar{f}_{i,{\scriptstyle \mathrm{trop}}}\right)^{2}}\right]^{\frac{n}{2}+1}}.\label{eq: strong tropical Gauss-Kronecker curvature}
\end{eqnarray}
Main difference between the two sectors comes from second derivatives
of (\ref{eq: non-ideal tropical mapping homogeneous degree d double scaling})

\begin{eqnarray}
\frac{\partial F_{\lambda}}{\partial x_{i}^{\star}\partial x_{j}^{\star}}&=\sum_{\alpha=1}^{m}w_{\alpha}(\lambda)\cdot\frac{\partial f_{\alpha}}{\partial x_{i}^{\star}\partial x_{j}^{\star}}\nonumber\\&+\lambda^{d}w_{\alpha}(\lambda)\cdot\left(\frac{\partial f_{\alpha}}{\partial x_{i}^{\star}}\frac{\partial f_{\alpha}}{\partial x_{j}^{\star}}-\sum_{\beta=1}^{m}w_{\beta}(\lambda)\cdot\frac{\partial f_{\beta}}{\partial x_{j}^{\star}}\frac{\partial f_{\alpha}}{\partial x_{i}^{\star}}\right).\label{eq: second derivatives slow}
\end{eqnarray}
Indeed, if $\alpha\neq\bar{\alpha}_{p}$ for all $p=1,\dots,r$ then
${\displaystyle \lim_{\lambda\rightarrow\infty}\lambda^{g}\cdot w_{\lambda,\alpha}}=0$
for all real $g$. Then, non-vanishing terms in (\ref{eq: second derivatives slow})
are of the form 
\begin{equation}
{\displaystyle \sum_{p=1}^{r}w_{\bar{\alpha}_{p}}(\lambda)\cdot\frac{\partial f_{\bar{\alpha}_{p}}}{\partial x_{i}^{\star}\partial x_{j}^{\star}}}\label{eq: 0-th order term second derivative slow degree d}
\end{equation}
 or 
\begin{equation}
{\displaystyle \sum_{p=1}^{r}\lambda^{d}w_{\bar{\alpha}_{p}}(\lambda)\cdot\left(\frac{\partial f_{\bar{\alpha}_{p}}}{\partial x_{i}^{\star}}\frac{\partial f_{\bar{\alpha}_{p}}}{\partial x_{j}^{\star}}-\sum_{q=1}^{r}w_{\bar{\alpha}_{q}}(\lambda)\cdot\frac{\partial f_{\bar{\alpha}_{q}}}{\partial x_{j}^{\star}}\frac{\partial f_{\bar{\alpha}_{p}}}{\partial x_{i}^{\star}}\right).}\label{eq: d-th order term second derivative slow degree d}
\end{equation}
If $r=1$ last term is ${\displaystyle \lambda^{d}w_{\bar{\alpha}_{1}}(\lambda)\cdot\left(\frac{\partial f_{\bar{\alpha}_{1}}}{\partial x_{i}^{\star}}\frac{\partial f_{\bar{\alpha}_{1}}}{\partial x_{j}^{\star}}-\frac{\partial f_{\bar{\alpha}_{1}}}{\partial x_{j}^{\star}}\frac{\partial f_{\bar{\alpha}_{1}}}{\partial x_{i}^{\star}}\right)=0}$
and (\ref{eq: 0-th order term second derivative slow degree d}) dominates
for large $\lambda$. In this case one gets formulae (\ref{eq: non-uniform, degree d tropical case, Christoffel})-(\ref{eq: non-uniform, degree d tropical case, Gauss-Kronecker}).
On the other hand, if $r>1$ then (\ref{eq: d-th order term second derivative slow degree d})
is non vanishing in general and this leads to expressions (\ref{eq: strong tropical Christoffel})-(\ref{eq: strong tropical Gauss-Kronecker curvature}).
Again, different patterns can be observed depending on the specific form of
interactions.

\end{document}